\documentclass[prl,showpacs,amsmath,preprint]{revtex4}

\usepackage{graphicx}
\usepackage[dvips]{color}

\usepackage[dvips]{color}

\newcommand{\BAN}{\ensuremath{B_{1g}\,}}
\newcommand{\BN}{\ensuremath{B_{2g}\,}}
\newcommand{\OmAN}{\ensuremath{\Omega_{B_{1g}}\,}}
\newcommand{\OmN}{\ensuremath{\Omega_{B_{2g}}\,}}
\newcommand\Deltam{\ensuremath{\Delta_{max}\,}}
\newcommand{\ZAN}{\ensuremath{Z\Lambda_{AN}\,}}
\newcommand{\ZN}{\ensuremath{Z\Lambda_{N}\,}}
\newcommand{\vD}{\ensuremath{v_\Delta\,}}
\newcommand{\phiN}{\ensuremath{\phi_{N}\,}}
\newcommand{\cxcy}{\ensuremath{\cos k_x-\cos k_y\,}}
\newcommand{\arc}{\ensuremath{f_c\,}}

\begin{document}

\title{Electronic Raman Scattering in copper oxide Superconductors: Understanding the Phase Diagram}

\author{A. Sacuto, Y. Gallais, M. Cazayous, S. Blanc and M.A.M\'easson}
\affiliation{Laboratoire Mat\'eriaux et Ph\'enom$\grave{e}$nes Quantiques (UMR 7162 CNRS),
Universit\'e Paris Diderot-Paris 7, Bat. Condorcet, 75205 Paris Cedex 13, France}

\author{J. S. Wen, Z. J. Xu and G. D. Gu}
\affiliation {Matter Physics and Materials Science, Brookhaven National Laboratory (BNL), Upton, NY 11973, USA}.

\author{D. Colson}
\affiliation{Service de Physique de l'Etat Condens\'{e}, CEA-Saclay, 91191 Gif-sur-Yvette, France}

\date{\today}

\begin{abstract}

Electronic Raman scattering measurements have been performed on hole doped copper oxide (cuprate) superconductors as a function of temperature and doping level. In the superconducting state coherent Bogoliubov quasiparticles develop preferentially over the nodal region in the underdoped regime. We can then define the fraction of coherent Fermi surface, $f_c$ around the nodes for which quasiparticles are well defined and superconductivity sets in. We find that $f_c$ is doping dependent and leads to the emergence of two energy scales. We then establish in a one single gap scenario, that the critical temperature $T_{c} \propto f_{c}\Delta_{max}$ where $\Delta_{max}$ is the maximum amplitude of the d-wave superconducting gap. In the normal state, the loss of antinodal quasiparticles spectral weight detected in the superconducting state persists and the spectral weight is only restored above the pseudogap temperature $T*$. Such a dichotomy in the quasiparticles dynamics of underdoped cuprates is responsible for the emergence of the two energy scales in the superconducting state and the appearance of the pseudogap in the normal state. We propose a 3D phase diagram where both the temperature and the energy phase diagrams have been plotted together. This 3D diagram advocates in favor of a low temperature phase transition inside the superconducting dome. We anticipate that the development of coherent excitations on a restricted part of the Fermi surface only is a general feature in high $T_c$ cuprate superconductors as the Mott insulating is approaching.

\par
--------------------------------------------------------------------------------------------------------------------------
\par

Des mesures de diffusion Raman \'{e}lectronique ont \'{e}t\'{e} men\'{e}es sur les oxydes de cuivre supraconducteurs dop\'{e}s en trous en fonction de la temp\'{e}rature et du dopage. Dans l'\'{e}tat supraconducteur du r\'{e}gime sous dop\'{e} le poids spectral des quasiparticules de Bogoliubov reste important dans les regions nodales alors qu'il est r\'{e}duit dans les regions antinodales. On peut alors d\'{e}finir la fraction coh\'{e}rente de la surface de Fermi, $f_c$, autour des noeuds, sur laquelle la supraconductivit\'{e} se d\'{e}veloppe. Nous avons d\'{e}couvert que $f_c$ d\'{e}pend du dopage et est \`a l'origine de l'apparition de deux \'{e}chelles d'\'{e}nergie dans le r\'{e}gime sous dop\'{e} de l'\'{e}tat supraconducteur. Nous avons alors \'{e}tabli dans un scenario \`a  un seul gap que la temp\'{e}rature critique $T_{c} \propto f_{c}\Delta_{max}$ o\`u $\Delta_{max}$ est l'amplitude maximale d'un gap de sym\'{e}trie d. Dans l'\'{e}tat normal, la perte de poids spectral (observ\'{e}e dans l'\'{e}tat supraconducteur) presiste et ne disparait qu'au dessus de la temp\'{e}rature de pseudogap $T*$. Nous pensons que cette forte dichotomie dans la dynamique des quasiparticules est responsable \`a la fois de l'apparition des deux \'{e}chelles d'\'{e}nergie dans l'\'{e}tat supraconducteur et du pseudogap dans l'\'{e}tat normal des cuprates sous dop\'{e}s. Nous proposons un diagramme de phase 3D o\`u  sont repr\'{e}sent\'{e}s simultan\'{e}ement les diagrammes de phase en temp\'{e}rature et en \'{e}nergie. Ce diagramme 3D privil\'{e}gie un scenario o\`u un changement d'\'{e}tat devrait exister \`a  basse temp\'{e}rature \`a l'interieur du dome supraconducteur. Nous pensons que le d\'{e}veloppement des excitations coh\'{e}rentes sur des portions restreintes de la surface de Fermi est un trait caract\'{e}ristique des cuprates \`a haute temp\'{e}rature \`a l'approche d'un isolant de Mott. 

\end{abstract}

\maketitle

\section*{INTRODUCTION}

One of the most challenging issues in copper oxide superconductors is to understand how superconductivity emerges from a Mott insulating state as a hole concentration (doping level,~\textit{p}) is increased \cite{Anderson}. Cuprates consist of an alternated stacking of $CuO_2$ planes and reservoir planes . The low energy electronic structure of these planes is characterized by a single energy band \cite{Pickett}. At low doping level, this energy band is half filled. Band theory would predict this to be a metal but the actual material is an insulator. The origin of this insulating behavior is the Coulomb repulsion which prevent the hopping of an electron from one Cu site to the next. The electrons are then localized. The spins of these Cu ions form an antiferromagnetic order (known as a N\'eel lattice). As the doping level increases, electrons are transfered from $CuO_2$ planes to the ``reservoir'' planes. Holes then appear in the $CuO_2$ planes allowing the electron hopping from one Cu site to an other, and so rapidly destroy the N\'eel lattice. En artificial metal is then built and remarkably, around $p\approx0.05$, a superconducting state emerges (see fig.~1-a). 

\begin{figure}[!ht]
\begin{center}
\includegraphics[width=8cm]{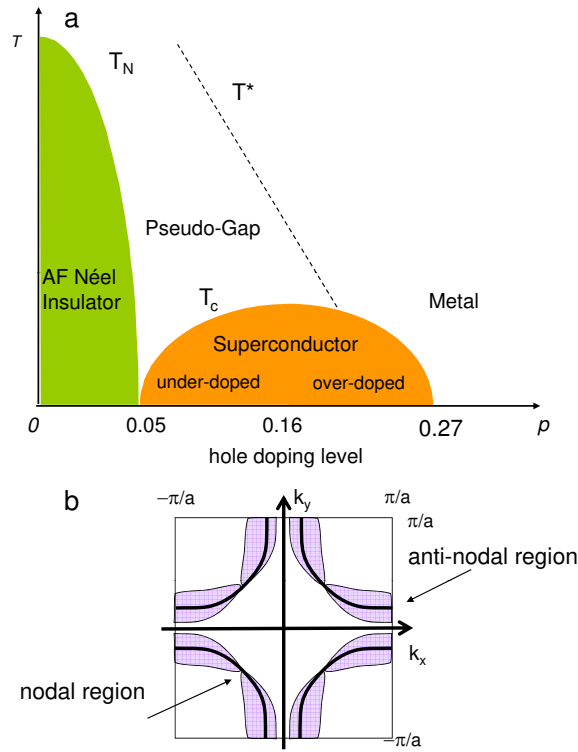}
\end{center}\vspace{-7mm}
\caption{(a) cuprate phase diagram (b) amplitude of a supercondcuting d-wave gap in the first Brillouin zone.} 
\label{fig1}
\end{figure}

The evolution of the superconducting transition temperature, $T_c$, as a function of the hole doping level, is remarkably universal. $T_c$ has a dome-like shape and exhibits two distinct regimes : (i) the underdoped regime where the critical temperature $T_c$ increases with doping until the optimal doping, $p\approx0.16$ and (ii) the overdoped regime where $T_c$ decreases with doping and vanishes for  $p\approx0.27$. It is now established that the superconducting gap has a dominant d-wave symmetry across the entire phase diagram although a smaller s-wave component cannot be ruled out \cite{Tsuei,Khasanov}. The superconducting gap reaches its maximum values along the principal axes of the Brillouin zone (BZ) and vanishes along the diagonal of the BZ. This corresponds respectively to the antinodal and nodal directions (see fig.~1b).

In the underdoped regime, above $T_{c}$ and below the temperature $T*$ a pseudogap state develops. It corresponds to a partial suppression of spin and charge excitations \cite{Alloul,Timusk} and it is also associated to broken symmetries \cite{Fauque,Xia,Lawler,He}.

On one hand, a growing number of transport measurements such as electrical and thermal conductivities, entropy, heat capacity and Hall coefficient, see for a review \cite{Tallon,Taillefer,Alloul2} advocate in favor of a temperature phase diagram where $T*$ do not merge with $T_c$ in the overdoped side but rather cuts through (or end at) the $T_c$ dome (see fig.~2a). 

\begin{figure}[!ht]
\begin{center}
\includegraphics[width=10cm]{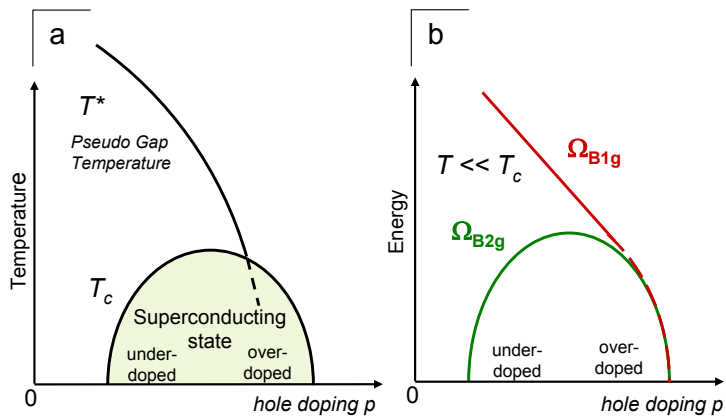}
\end{center}\vspace{-7mm}
\caption{(a) temperature and (b) energy phase diagrams versus hole doping level} 
\label{fig2}
\end{figure}

On the other hand, spectroscopic investigations such as Andreev-Saint-James reflection \cite{Deutscher}, electronic Raman sacttering (ERS) \cite{LeTacon,Guyard01,Gallais,Opel} angle resolved photoemission spectroscopy (ARPES)\cite{Mesot,Tanaka,Lee}, infrared reflectivty (IR)\cite{Bernhard},  and scanning tunneling microscopy (STM) \cite{Boyer,Kohsaka,Yazdani} lead to an energy phase diagram at low temperature (well below $T_{c}$) where a single energy scale is detected in the overdoped regime while two distinct energy scales appear in the underdoped side of the superconducting state (see fig.~2b). The low energy scale decreases while the high energy one increases with underdoping.  
How can we understand these two distinct phase diagrams in a global picture of high $T_c$ cuprate superconductors? Are we able to depict a 3D cuprate phase diagram which involves both energy and temperature as a function of hole doping level? 

These questions are a real challenge that we propose to address here. More precisely, we will try to reveal physics which control $T_c$ and $T*$ by performing and discussing Raman experiments on several high $T_c$ cuprate superconductors through a large range of doping levels and temperatures. This leads us to point out the emergence of two distinct quasiparticle dynamics and two distinct energy scales in the superconducting state of underdoped cuprates.

We show that Coherent Bogoliubov quasiparticles develop preferentially over a restricted region of the momentum-space in underdoped regime: around the nodal direction. The density of Cooper pairs appears to be strongly anisotropic in momentum space with underdoping. Most of the supercurrent is then carried out by electronic states around the nodal region in the momentum space. This contrasts to conventional superconductors where superconductivity develops uniformly along the normal-state Fermi surface. 

We can then define the fraction of coherent Fermi surface, $f_c$ around the nodes for which quasiparticles are well defined and superconductivity sets in. We find that $f_c$ is doping dependent and we establish that  $T_{c} \propto f_{c}\Delta_{max}$ where $\Delta_{max}$ is the maximum amplitude of the d-wave superconducting gap. This new relation differs from the standard BCS theory and give us some clues for increasing $T_c$ in the cuprates. 

Just above $T_c$, in the underdoped regime, the fraction of coherent Fermi surface is still observable. This manifests experimentally by a sizeable quasiparticle spectral weight in the nodal region while it is strongly reduced in the antinodal region. This is the signature of the pseudogap state. The quasiparticle spectral weight in the antinodal region is only recovered above $T*$. 

The loss of coherent quasiparticles in the antinodal region is then acting as a foe of superconductivity since it prevents from the formation of coherent Cooper pairs around the antinodes in underdoped cuprates. Loss of coherent quasiparticles on restricted parts of the Fermi surface is then responsible for both the existence of the two energy scales in the superconducting state and the appearance of the pseudogap in the normal state.  

These results have bearing on the fundamental problem of how superconductivity emerges as holes are doped into a Mott insulating state. We anticipate that the development of coherent excitations on a restricted part of the Fermi surface only is a general feature in high $T_c$ cuprate superconductors.

\section*{ELECTRONIC RAMAN SCATTERING}
\par

Raman scattering is usually known for its ability to probe the vibrational modes of the crystal lattice. However Raman scattering is also an efficient tool to investigate electronic excitations in the spin and charge channels such as collective modes (magnon, plasmon) or single particles excitations (quasiparticles). Indeed, electronic Raman scattering (ERS), like angle resolved photoemission spectroscopy (ARPES), is both an energy and a momentum probe of quasiparticles. ERS allows us to reach an energy accuracy of less than one tenth meV on a selected part of the BZ. ERS is an inelastic light scattering process where an incident photon is absorbed by the crystal and a scattered one is emitted, with the simultaneous creation (Stokes) or annihilation (anti-Stokes) of an electronic excitation. ERS is particulary suitable for cuprates where the light penetration depth is typically of the order of $100~nm$ corresponding roughly to one hundred cells iradiated. Here we will deal only with Stokes process illustrated in fig.~3.

\begin{figure}[!ht]
\begin{center}
\includegraphics[width=10cm]{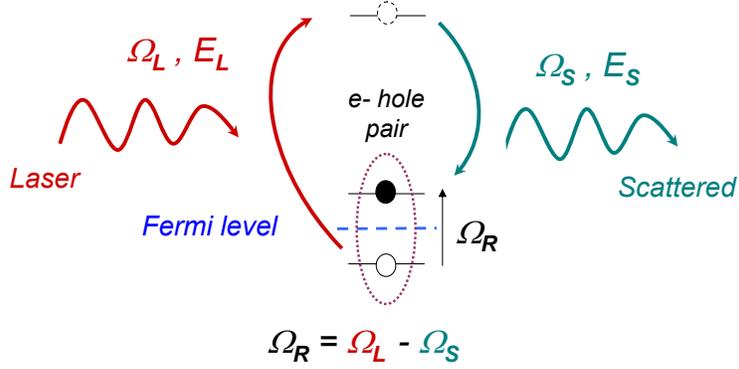}
\end{center}\vspace{-7mm}
\caption{Electronic Raman scattering process (Stokes process): a crystal is irradiated by a monochromatic wavelength of a laser beam and the scattered light is collected. The difference in frequency between the laser and scattered frequencies ($\Omega_L$ and $\Omega_S$ respectively) is called the Raman shift, $\Omega_R$  and corresponds to the energy of an electron-hole pair excitation around the Fermi level.}
\label{fig3}
\end{figure}

Since two photons are coming into play, Raman scattering is a second order process in the electromagnetic field. This second order effective interaction with electronic excitations comes from both a direct second order term in the interaction Hamiltonian, and from a first order term treated up to second order in perturbation. As introduced by Abrikosov and Genkin \cite{Abrikosov}, one may consider that both terms can be gathered in a single effective second order term in the Hamiltonian which can be written as:
\begin{eqnarray}
{H}_{R}=\frac{{e}^{2}}{m} \left\langle{{A}_{S}{A}_{L}}
\right\rangle{e}^{-i\Omega t}{\hat{\rho }}_{q}
\label{eq1}
\end{eqnarray}
where $e$ is the electronic charge and  $m$  its bare mass. $ A_{L}$ and $ A_{S}$ are the vector potentials of the incoming laser and scattered light, and the bracket is for the proper matrix element over the photons states. The difference between the incident and the scattered photon frequencies is noted $\Omega = \Omega_{L} - \Omega_{S} $ , and the difference between the photon momenta is $ q = {k}_{L} -{k}_{S} $. The operator ${\hat{\rho }}_{q}$ is given by :
\begin{eqnarray}
{\hat{\rho }}_{q}=\sum\nolimits\limits_{{n}_{f},{n}_{i},k}^{}
{\rm \gamma }_{{n}_{f},{n}_{i},k}
{\rm c}_{{n}_{f},k\rm +q}^{\rm +}{c}_{{n}_{i},k}
\label{eq2a}
\end{eqnarray}

It is quite similar to the standard density operator, where $k$ is the initial electronic momentum and  $n_{f}, n_{i}$ are the final and inital electronic bands. The only difference is the term ${\gamma }_{{n}_{f},{n}_{i},\bf k}$ for the scattering process, known as the Raman vertex which is given explicitely by :
\begin{eqnarray}
\lefteqn{{\gamma }_{{n}_{f},{n}_{i},\bf k}\rm
={e^{*}}_{\rm S}.{e}_{\rm L}{\bf
\delta }_{{\rm n}_{f},{n}_{i}}}  \nonumber \\
\nonumber  \\ [.1cm]
 & & + \frac{1} {\hbar m}\sum\nolimits\limits_{{n}_{m}}
\frac{\left\langle{{\rm n}_{f},k\rm +q\rm
 \left|{{e}^{-i{k}_{\rm s}.r}\rm {e^{*}}_{\rm
S}.}\right|{\rm n}_{m},k\rm +{
k}_{L}}\right\rangle\left\langle{{\rm n}_{m},k\rm
+{ k}_{L}\rm \left|{{e}^{i{k}_{\rm L}.
r}\rm {e}_{\rm L}.p}\right|{\rm n}_{i},
k}\right\rangle} {{\rm \varepsilon }_{{n}_{i},
k}\rm -{\varepsilon }_{{n}_{m},k\rm +{k}_{L}}\rm
+{\Omega }_{L}+i\eta } \nonumber \\
+(L\leftrightarrow S)
\label{eq2b}
\end{eqnarray}

${e^{*}}_{\rm S}$ and ${e}_{\rm L}$ are respectively the electric field polarizations of the incident and scattered light and  ${\rm \varepsilon }_{{n},k}\rm$ the electronic states.

The Raman vertex depends on the electronic band structure of the material studied and it is far to be easy to calculate it explicitly. Stricly speaking, the Raman vertex depends on $k,q,\Omega_{L}$ and $\Omega$. However, in the visible range which is the applied field of ERS, the photon momentum transfered $q$ is negligeable with respect to the BZ. Pratically, we consider $q=0$ in the Raman scattering process. Moreover for low frequency range (typically $\Omega$ less than 1/8 eV $\approx 1000~cm^{-1}$) in comparison with the electronic transitions (2 eV),  we consider that the Raman vertex does not depend on $\Omega$ \cite{Sacuto}. The Raman vertex is $k$ dependent however and its contraction in eq.~3, by the incident and scattered electric fields allows us to select different part of the BZ. Indeed, cuprates have a pure or slightly distored tetragonal structure. As a consequence, the Raman vertex (related to $CuO_2$ plane) can  be decomposed on the basis of the 2D irreductible representations of the $D_{4h}$ space group . 

\begin{eqnarray}
\tilde{\gamma}(k) = \gamma_{A_{1g}}(k) \begin{pmatrix} 1&0 \\ 0&1 \end{pmatrix} + \gamma_{B_{1g}}(k) \begin{pmatrix} 1&0 \\ 0&-1 \end{pmatrix}+ \gamma_{B_{2g}}(k) \begin{pmatrix} 0&1 \\ 1&0 \end{pmatrix}
\end{eqnarray}

In a such a case, the contraction of the Raman vertex tensor ($\vec{e}^{*}_{S}.\tilde{\gamma}(k).\vec{e}_{L}$) by the incident and scattered electric field polarizations fix one of its component, each component having a well defined symmetry which corresponds to a specific  momentum space dependence.

\begin{figure}[!ht]
\begin{center}
\includegraphics[width=10cm]{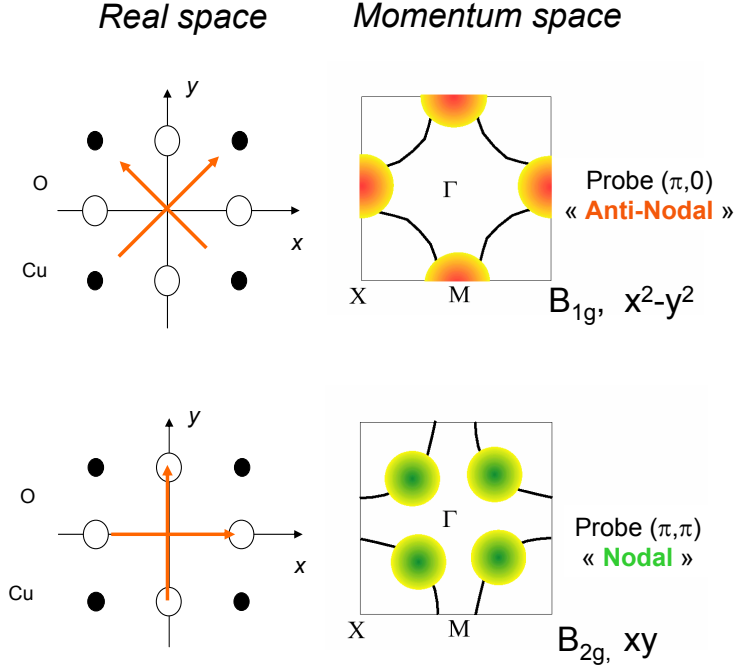}
\end{center}\vspace{-7mm}
\caption{Raman selection rules in cuprates.  The cross polarizations at $45^\circ$ from the copper oxide bonds in the real space will probe the principal axes of the BZ in the momentum space ($B_{1g}$ symmetry). The cross polarizations along the copper-oxide bonds in the real space will probe the diagonal of the BZ in the momentum space ($B_{2g}$ symmetry). The antinodal (AN) and nodal (N) regions refer to the $d-wave$ superconducting gap symmetry which takes its maximum amplitude along the principal axes and vanishes along the diagonals of the Brillouin zone.}
\label{fig4}
\end{figure}

As for an example, see fig.~4, diagonal cross polarizations $\begin{pmatrix} 1\\1 \end{pmatrix}$ and $\begin{pmatrix} 1\\-1 \end{pmatrix}$

 active the \BAN tensor component  while the \BN and $A_{1g}$ tensor components are not active. The \BAN tensor (with respect to the Neuman's theorem) \cite{Burns} transforms as (${k_x}^2-{k_y}^2$). It vanishes along the diagonal of the BZ and therefore probes mainly the principal axes of the BZ (the antinodes). On the opposite, cross polarizations $\begin{pmatrix} 1\\0 \end{pmatrix}$ and $\begin{pmatrix} 0\\1 \end{pmatrix}$ will select the \BN tensor component which transforms as ($k_{x}k_{y}$) and probes mainly the diagonal of the BZ (the nodes). 

In summary, by a judicious choice of the incident and scattering electric fields we are able to probe different parts of the BZ. The contraction of the Raman vertex acts as a filter which hides some specific regions of the BZ. In cuprates, we can probe the nodal region (N) and the antinodal regions (AN). 

Raman experiments give a direct access to the Fourier transform of the density-density correlation function called the ``dynamical structure factor'' \cite{Klein}:

\begin{eqnarray}
S(q,\Omega,T)= \int\frac{dt}{2\pi}e^{i\Omega t}\left\langle{{\hat{\rho }}^{\dagger}(q,t){\hat{\rho }}(q,0)}\right\rangle_{T}
\label{eq3}
\end{eqnarray}
where $\left\langle ..\right\rangle_{T}$ is the thermal average. 

According to the fluctuation dissipation theorem \cite{Callen}, $S(q,\Omega,T)$ is related to the imaginary part $\chi''(q,\Omega,T)$ of the response function $\chi(q,\Omega,T)$ as follows: 
\begin{eqnarray}
S(q,\Omega,T)=\frac{\hbar}{\pi}(1+n(\Omega,T))\chi''(q,\Omega,T)
\label{eq4}
\end{eqnarray}

where $n(\Omega,T) =(e^{-\hbar\Omega/k_{B}T}-1)^{-1}$ is the Bose-Einstein factor.

$\chi''(q,\Omega,T)$ is related to the electronic density fluctuations induced by the electric field of the incident light into the crystal. 

The Raman response function (or the dynamical structure factor) can be explicitly calculated in some specific cases such as the normal state of a Fermi liquid with or wihout impurity or the superconducting state in BCS theory. We can consider several ways to calculate the dynamical structure factor, one of them consists to use the Matsubara formalism which leads to an analytical expression of the Raman response function at finite temperature. This formalism will be used in the next sessions. We invite the reader to refer to the references \cite{Rychkasen,Mahan,Devereaux1} for more details.

\section*{K-SPACE ISLANDS OF COHERENT COOPER PAIRS IN THE SUPERCONDUCTING STATE OF UNDERDOPED CUPRATES}

We have first performed Raman measurements on a single $CuO_2$ layer compound: $HgBa_{2}CuO_{4+\delta }$ ($Hg-1201$). Raman spectra in both \BAN (AN) and \BN (N) geometries (for several doping levels) are displayed in fig.~5. 

\begin{figure}[!ht]
\begin{center}
\includegraphics[width=15cm]{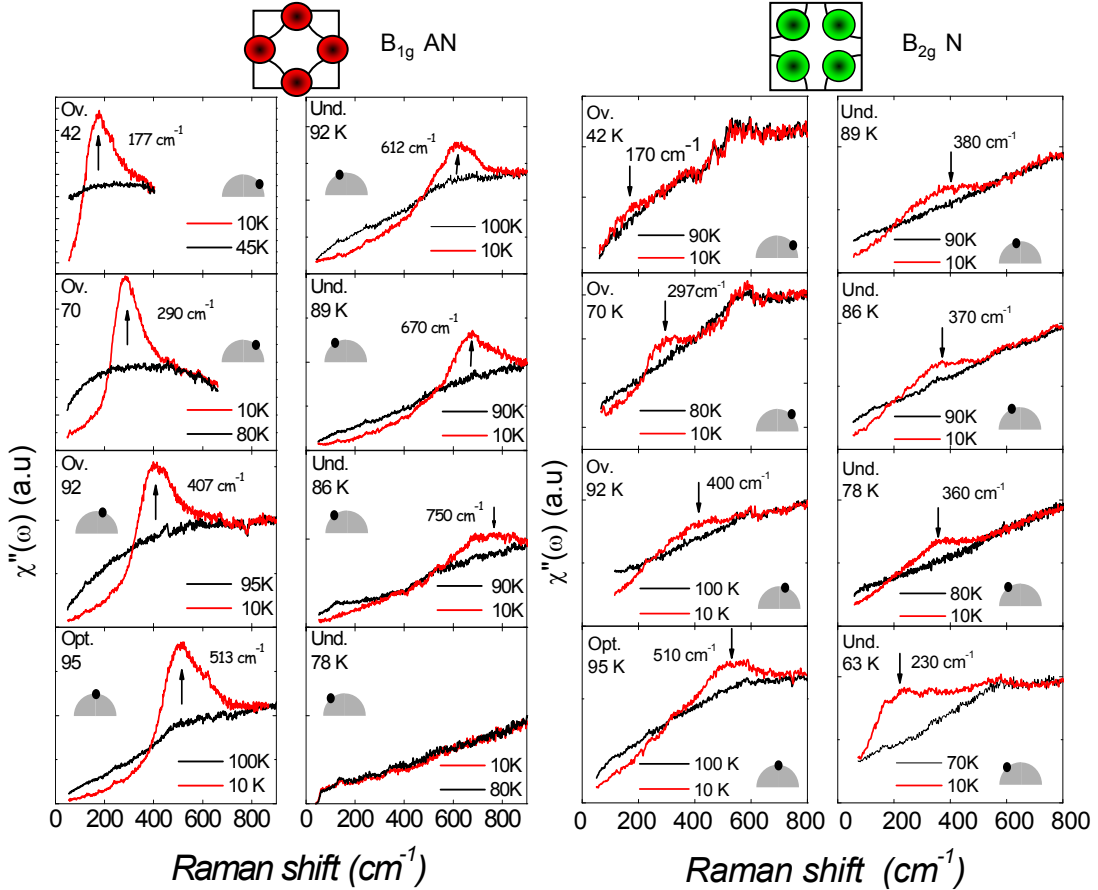}
\end{center}\vspace{-8mm}
\caption{Raman spectra of $HgBa_{2}CuO_{4+\delta}$($Hg-1201$) single crystals for several doping levels. The gray (red) curves correspond to the Raman spectra of the superconducting state in the \BAN (AN) and \BN (N) geometries respectively. The black curves correspond to the Raman spectra in the normal state just above $T_c$ ~\cite{LeTacon}.}
\label{fig5}.
\end{figure}

At a first glance, we observe in the superconducting state of overdoped samples a strong peak in the \BAN geometry and a weaker one in the \BN geometry (first and third pannels from the left). Remarquably these \BAN and \BN peak energies exhibit two distinct doping dependence in the underdoped regime (second and forth pannels from the left). The \BAN peak increases in energy while the \BN one decreases in energy with underdoping.

At the optimal doping level, ($T_c$ = 95~K, bottom of the first and third pannels from the left), the \BAN low energy spectrum (below $400~cm^{-1}$) exhibits a nearly cubic frequency dependence. On the opposite, the low energy \BN spectrum displays a linear frequency dependence. These two distinct power laws (cubic and linear) are the Raman signature of a d-wave superconducting gap and can be qualitatively understood as follows (see for more details \cite{Devereaux2}). In the \BAN geometry (around the antinodes, see fig.~6-a), the superconducting gap has its maximum amplitude which prevents low energy electronic excitations. We then expect the electronic continuum to be weak  below $2\Delta_{0}$, (see fig.~6-b).

On the opposite, in \BN geometry (close to the nodes, see fig.~6-c), the amplitude of the superconducting gap vanishes. This allows substantial low energy electronic excitations. Since the number of available electronic states increases with energy, we expect the Raman spectrum to exhibit a linear frequency dependence at low energy (see fig.~6-d). 

\begin{figure}[!ht]
\begin{center}
\includegraphics[width=10cm]{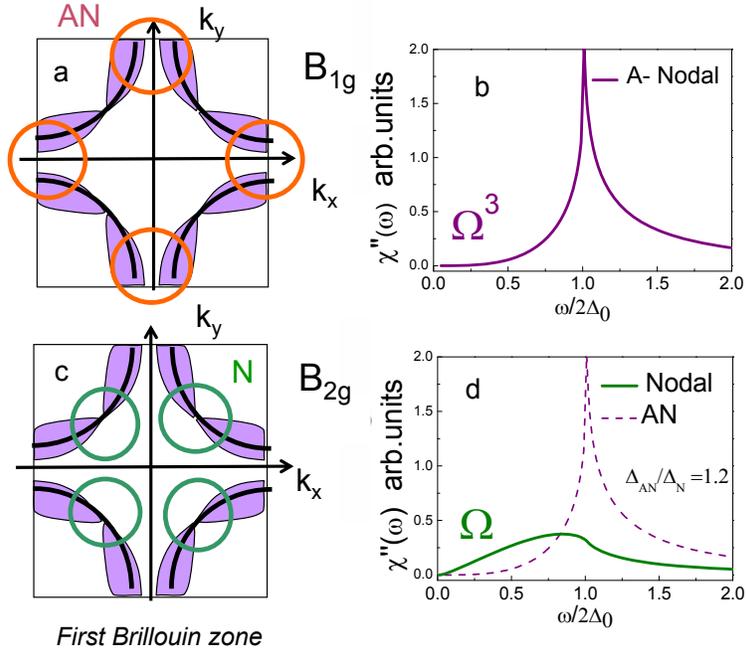}
\end{center}\vspace{-8mm}
\caption{(a) Antinodal regions probed by the $B_{1g}$ geometry, (gray (orange) circles) and (b) the calculated \BAN Raman response for a d-wave superconducting gap. (c) Nodal regions probed by the $B_{2g}$ geometry, (gray (green) circles) and (d) the calculated \BN Raman response for a d-wave gap. The (violet) petals correspond to the amplitude of the d-wave superconducting gap. It takes its maximum amplitude along the principal axes of the BZ and vanishes along the diagonals.}
\label{fig6}.
\end{figure}

\subsection*{Area of the nodal and antinodal superconducting peaks and evolution of the density of Cooper pairs with underdoping}

Beyond the analysis of the low energy electronic continuum in  \BAN and \BN geometries, fig.~5 reveals that the antinodal \BAN peak measured at $T=10~K$ exhibits a strong decrease in intensity with underdoping before disapearing (close to $p=0.12$, $T_c =78~K$) while the nodal \BN peak persists down to the lowest doping level ($p=0.09$, $T_c=63~K$).

In order to make reliable comparison between the Raman intensities of the \BAN and \BN superconducting peaks, we have performed quantitative Raman measurements which allow a direct comparison of the intensities for different doping levels (see Methods).  

Obtaining intrinsic Raman measurements on cuprates with various doping levels is a true challenge for experimentalists. It requires not only an extremely high level of control of the crystal surface quality, the optical set up but also the knowledge of the optical constants for each crystal studied. In order to overcome these difficulties, we have chosen to work on $Bi_{2}Sr_{2}CaCu_{2}O_{8+d}$ ($Bi-2212$) system rather than on the $Hg-1201$ one \cite{LeTacon,Guyard01,Gallais}. One of the reason is that $Bi-2212$ crystals can be easily cleaved providing large homogeneous surfaces ($\approx mm^{2}$) 

We have performed all the measurements during the same run and the crystals with various doping levels have been mounted on the same sample holder in order to keep the same optical configuration. With a laser spot of about $50~\mu m$ in diameter, we have measured Raman intensity variations of less than $5\%$ from one point to another on the same cleaved surface. Crucially, we have also observed only weak intensity changes for two distinct crystals of the same nominal doping level mounted side by side on the sample holder of the cryostat. These observations give us confidence that the doping dependence of the Raman intensity variations reported here are intrinsic. Finally, the Raman cross-section at each doping level was obtained by correcting the Raman response function for the optical constants \cite{Blanc}.

\begin{figure}[!ht]
\begin{center}
\includegraphics[width=15cm]{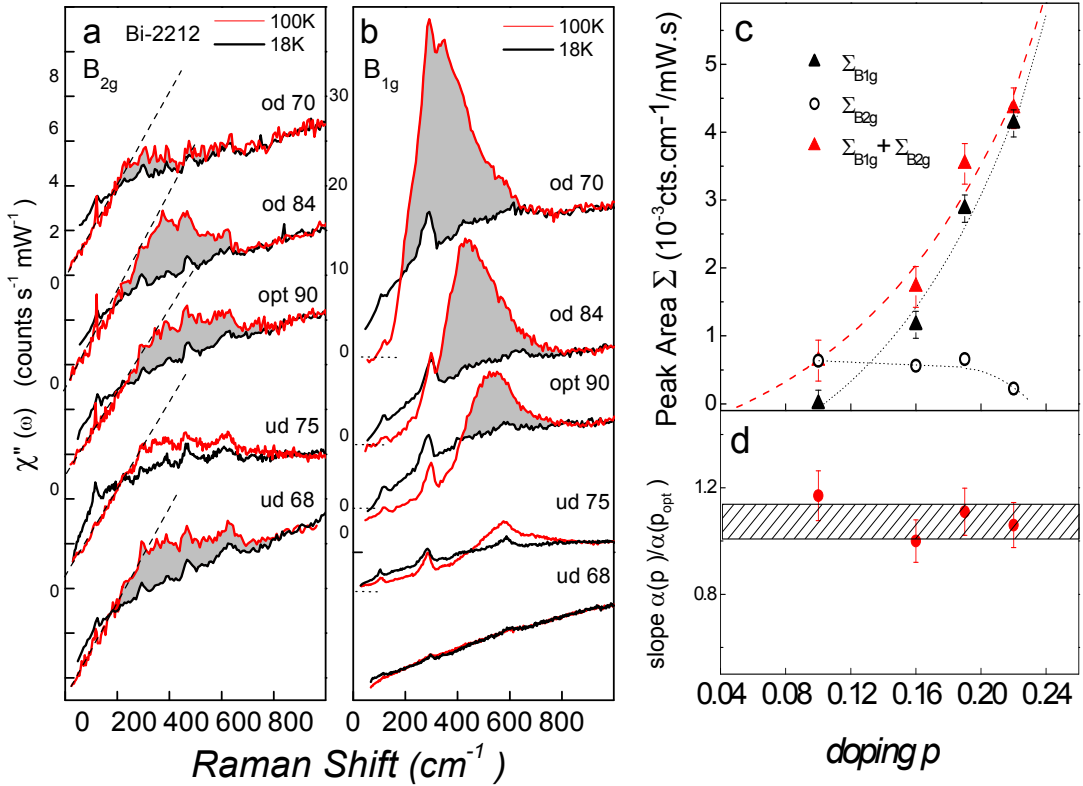}
\end{center}\vspace{-7mm}
\caption{Intrinsic Raman measurements on the superconducting and normal states in (a) $B_{2g}$ and (b) $B_{1g}$ geometries for distinct doping levels. The grey zones correspond to the subtraction between the superconducting and the normal Raman responses under the superconducting pair breaking peaks. (c)$B_{1g}$ and $B_{2g}$ superconducting peak areas ($\Sigma_{B_{1g}},\Sigma_{B_{2g}}$). (d) slope of the $B_{2g}$ electronic continuum \cite{Blanc}.}
\label{fig7}.
\end{figure}

Figures~7-a and b display the intrinsic \BN and \BAN superconducting responses of $Bi-2212$ for several doping levels in the superconducting and normal states. 

We focus first on the \BAN and \BN peak areas deduced from the subtraction between the superconducting and the normal Raman responses (in gray in fig.~7a and b). 
 
Our data reveal a strong decrease of the area under the \BAN peak with underdoping. It disappears close to $p= 0.1$ while the \BN superconducting peak area slightly increases from $p=0.22$ to $0.19$ and then remains almost constant as the doping level is reduced down to $0.1$. The doping evolution of the \BN and \BAN peak area ($\Sigma_{B_{1g}},\Sigma_{B_{2g}}$) are reported on fig.~7-c. 

Are these two peaks truly associated to coherent excitations of the superconducting state or can subsist above $T_c$? In order to precisely answer to this question, we have performed Raman measurements as a function of both doping level and temperature.
In figure~8-a and b are displayed the Raman responses $\chi''(\Omega,T)$ of $Bi_{2}Sr_{2}CaCu_{2}O_{8+d}$ ($Bi-2212$) single crystals with different doping levels, in $B_{1g}$ (antinodal) and $B_{2g}$ (nodal) geometries, for several temperatures ranging from well below $T_c$ to $10$~K above $T_c$.

\begin{figure}[!ht]
\begin{center}
\includegraphics[width=15cm]{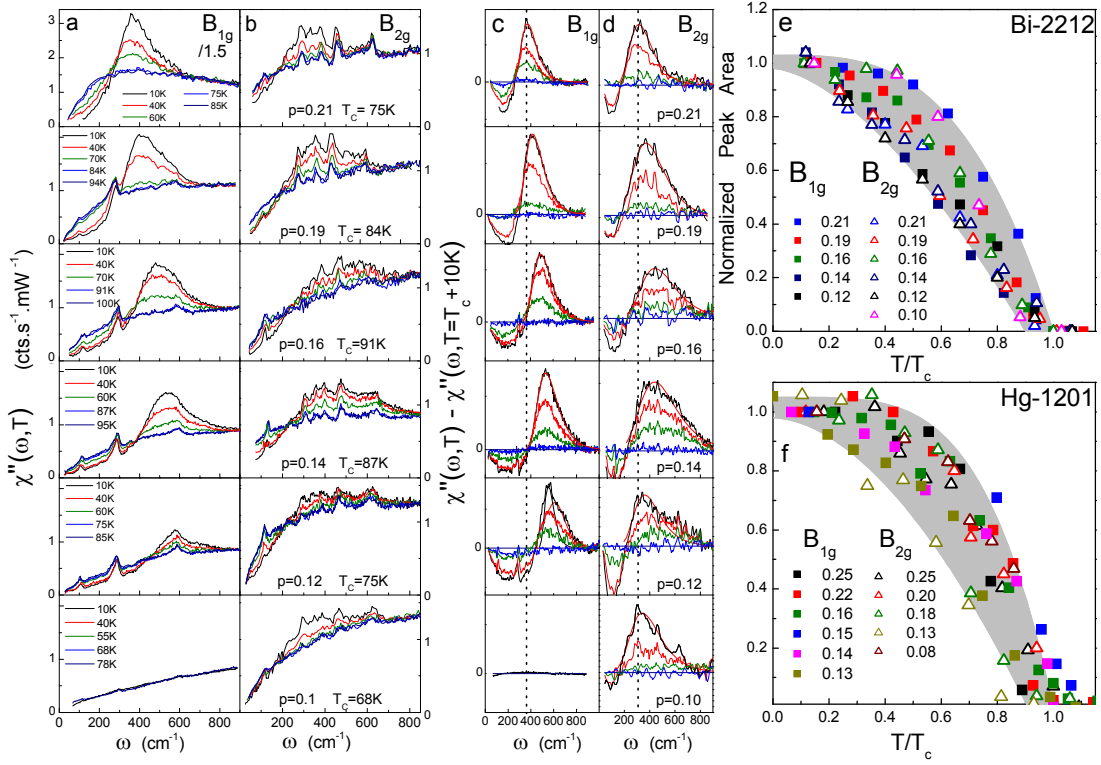}
\end{center}\vspace{-7mm}
\caption{Temperature dependence of the Raman spectra.
(a) and (b): Raman response, $\chi''(\omega,T)$ of $Bi-2212$ single crystals for several doping levels in $B_{1g}$ (AN) and $B_{2g}$ (N) geometries.
(c) and (d): Raman spectra subtracted from the one measured at $10~K$ above $T_c$ for each sample in each geometry.  A direct visual comparison of the subtracted spectra to a reference energy (chosen as the peak position for the most overdoped sample, drawn as a guide to the eyes) clearly
reveals the distinct doping-dependence of these two energy scales. (e) and (f): Temperature dependence of the normalized areas of the $B_{1g}$ and $B_{2g}$ peaks with respect to the area measured at $T=10~K$ for $Bi-2212$ and $Hg-1201$ crystals repectively \cite{Blanc2}.} 
\label{fig8}
\end{figure}

In both geometries, these spectra show the gradual decreasing of a peak (or rather a broad shoulder in the \BN geometry) as the sample is heated up to $T_c$. 

In order to clearly reveal the temperature-dependence of the \BAN and \BN peak areas,
we have plotted  in fig.~8-c-d, the difference between these spectra and the spectrum measured at $10$~K above $T_c$ on the same sample. From these subtracted spectra, we obtain the normalized areas of the \BAN and \BN peaks, which are displayed in fig.~8-e as a function of $T/T_c$ . Figure~8-f shows the \BAN and \BN areas for the $Hg-1201$ crystals. 

These plots demonstrate that the peak intensities vanish continuously at $T_c$, providing quantitative support to our interpretation as coherence peaks of the superconducting state.

What is the meaning of the superconducting peak area? For a non interacting Fermi liquid, in the framework of BCS theory, the Raman response in the limit $q\rightarrow0$ and use of Matsubara formalism \cite{Devereaux1,Klein2} leads to : 

 \begin{eqnarray}
\chi^{,,}_{\mu}(q=0,\Omega,T) =\pi  \sum_{k}(\gamma^{\mu}_{k})^{2}\tanh (\frac{E_{k}}{2k_{B}T})\frac{\left|\Delta_k\right|^{2}}{E_{k}^{2}}\delta(\Omega-2E_{k})
\label{eq5}
\end{eqnarray}

where $\mu $ refers to the $B_{1g}$ and $B_{2g}$ geometries, $ \gamma^{\mu}_{k}$ is the Raman vertex, $\Delta_k$, the superconducting gap and $k_B$ the Boltzman constant. $E_{k}$ is the Bogoliubov quasiparticle energy defined such as: $E_{k}=\sqrt{\xi_{k}^{2}+ \Delta^{2}_{k}}$ and  $\xi_{k}=\epsilon_{k}-\mu$.  $\epsilon_{k}$ is the electronic state energy and $\mu$ the chemical potential.
     
It is then straightforward to show that the integral of the Raman response over $\Omega$ when $T$ tends to zero, gives: 

\begin{eqnarray}          
\int{\chi^{,,}_{\mu}(\Omega)d\Omega}=\pi  \sum_{k}(\gamma^{\mu}_{k})^{2}\frac{\left|\Delta_k\right|^{2}}{E_{k}^{2}}
 \label{eq6}
\end{eqnarray}           
          
The sum $\sum_{k}\frac{\left|\Delta_k\right|^{2}}{E_{k}^{2}}$ is equal to $4\sum_{k}(u_kv_k)^2$ where $v_k^2$ and $u_k^2$ are the probabilities of the pair $(k\uparrow,-k\downarrow)$ being occupied and unoccupied respectively. This sum is non-vanishing only around the Fermi energy $E_F$ in the range of $2\Delta_k$ \cite{deGennes}. This quantity corresponds to the density of Cooper pairs, formed around the Fermi level as the gap is opening \cite{Leggett}. A priori, the density of coherent Cooper pairs is distinct from the superfluid density which is just the total carrier density at $T=0~K$. The integral of the Raman response is then proportional to the density of Cooper pairs, weighted by the square of the Raman vertex which selects specific area of the Brillouin zone: the nodal or the antinodal regions.

Applying this analysis to our data reveals that the superconducting peak area (in gray in fig.~7-a and b) provides a direct estimate of the density of Cooper pairs in the nodal and antinodal regions \cite{Note}.

The data reported in fig.~7-c show that the density of Cooper pairs is strongly anisotropic in the $k-$space as a function of doping level.

At low doping level, the density of Cooper pairs becomes very weak at the antinodes and vanishes below $p=0.1$, while it is still sizeable around the nodes. Therefore we are led to conclude that Cooper pairs are k-space confined. At low doping level Cooper pairs  form k-space islands around the nodes. This is consistent with the picture where most of the supercurrent is carried out by electrons' small patches centered on the nodal points on the underdoped regime as proposed by Ioffe and Millis \cite{Ioffe_98}. This picture is consistent with the loss of antinodal quasiparticles coherence reported in tunneling ~\cite{Kohsaka,McElroy} and ARPES ~\cite{Feng,Ding,Vishik}. The doping evolution of the density of Cooper pairs is sketched in fig.~9-a-c.

\begin{figure}[!ht]
\begin{center}
\includegraphics[width=10cm]{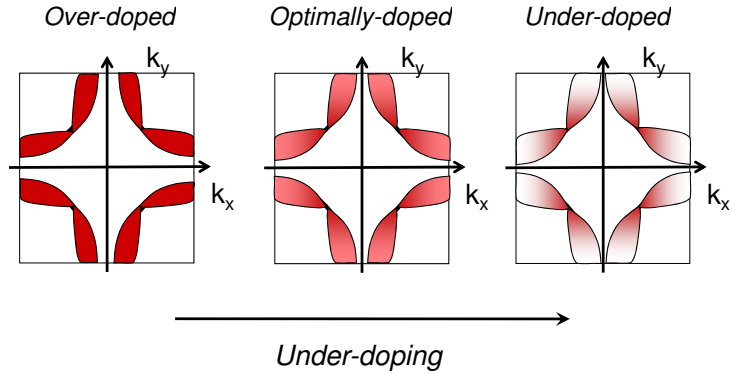}
\end{center}\vspace{-7mm}
\caption{Sketches of the d-wave superconducting gap amplitude in the momentum space for three distinct doping levels. The dark (red) zone corresponds to a high density of coherent Cooper pairs and the bright one to a low density of Cooper pairs. Cooper pairs developp preferentially around the diagonal of the Brillouin zone in the underdoped regime forming k-space island of Cooper pairs.}
\label{fig9}.
\end{figure}

\section*{TWO ENERGY SCALES IN THE SUPERCONDUCTING STATE OF UNDERDOPED CUPRATES} 

\subsection*{Nodal and Anti-Nodal energy scales in the underdoped side of the superconducting dome}

As pointed out in previous section (see fig.5), the \BAN and \BN peak energies exibit distinct doping dependences in both $Hg-1201$ and $Bi-2212$ systems. This can also be seen in fig.~8-c and d. The \BAN peak energy increases while the \BN one decreases with underdoping. This gives rise to two energy scales in the superconducting state of underdoped cuprates as reported in fig.~10-a. The \BAN and \BN energy scales coincide at high doping levels ($p\gtrsim 0.19$), but depart from each other as doping level is reduced. The \BAN energy scale increases monotonically as doping is reduced, while the \BN energy scale follows a dome-like shape approximately similar to that of the critical temperature $T_c$.

\begin{figure}[!ht]
\begin{center}
\includegraphics[width=9cm]{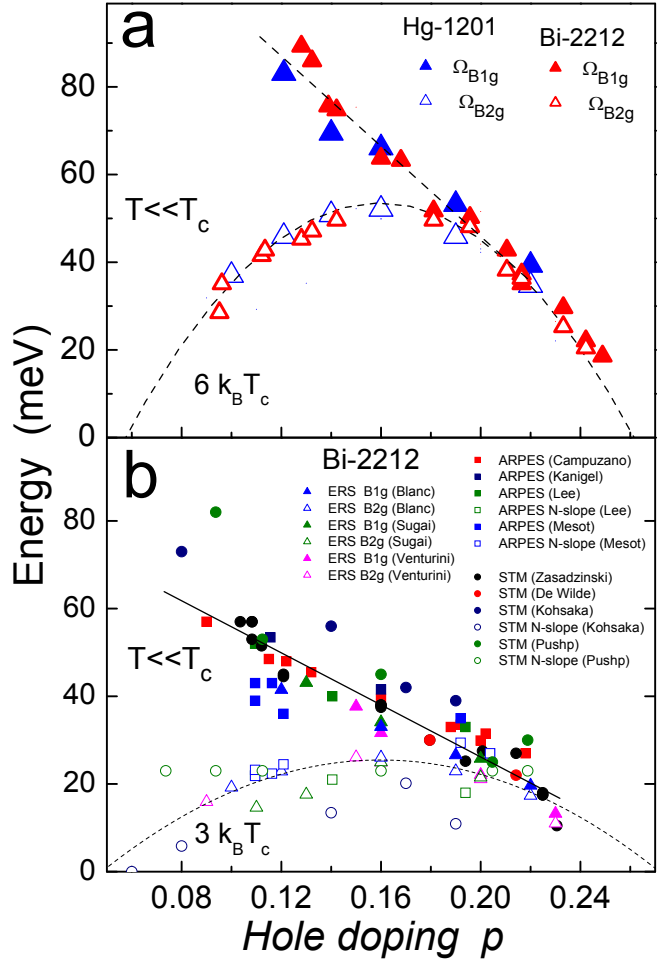}
\end{center}\vspace{-7mm}
\caption{The two energy scales in the superconducting state of underdoped cuprates, $T<<T_c$ ~\cite{LeTacon,Blanc}. (a) Doping dependence of the \BAN and \BN superconducting peak energies for both $Bi-2212$ and $Hg-1201$. (b) Plot of the two energy scales from several spectroscopic probes in $Bi-2212$ system well below $T_c$. To be compared to ARPES and STM data, ERS data have been divided by 2. The N-slope is the slope of the superconducting gap at the nodes deduced from ARPES and STM data. It is also quoted \vD.}
\label{fig10}
\end{figure}

In fig.~10-b, are reported on the same plot the ERS ~\cite{LeTacon,Venturini,Sugai}, tunneling \cite{Kohsaka,Yazdani,DeWilde,Zasadzinski} and ARPES \cite{Mesot,Lee,Campuzano,Kanigel} measurements performed on $Bi-2212$ system well below $T_c$. We clearly observe two energy scales as a function of doping level. The high energy scale (in filled symbols) corresponds to quasiparticles related to the antinodal region while the lower energy scale (in open symbols) corresponds to quasiparticle related to the nodal region.

\subsection*{Toward an understanding of the two energy scales in the underdoped side of the superconducting dome}
 
The origin and significance of these two scales are largely unexplained, although they have often been viewed as evidence for two distinct gaps in the superconducting state of under-doped cuprates. A popular view has been to associate one of the gap to the superconducting state while the other one is unrelated to superconductivity but associated with pseudogap~\cite{Tallon,Millis,Hufner}. In principle angular resolved photoemission spectroscopy (ARPES) should be able to distinguish between these two gaps but contradictory results have been reported up to now \cite{Terashima,Shi,Kondo,Chatterjee}. 

We have shown that the two energy scales disappear at $T_c$ and are associated with coherent excitations of the superconducting state (see fig.8). This leads us to another view.
Using a simple model, we show that these two energy scales do not require the existence of two distinct gaps: a pseudo gap and a superconducting one \cite{Storey,Belen,Leblanc,Tallon2}. Rather, a single d-wave superconducting gap with a loss of Bogoliubov quasiparticle spectral weight in the antinodal region is shown to reconcile spectroscopic and transport measurements in underdoped cuprates.

In order to shed light on the origin of the two energy scales revealed by Raman in \BAN and \BN geometries, we consider a very simple phenomenological model of a superconductor with a gap function $\Delta(\phi)$. 

The angle $\phi$ is associated with momentum k on the Fermi surface. The gap function vanishes at the nodal point $\Delta(\phi=\pi/4)=0$ while it is maximal at the antinodes $\Delta(\phi=0)=\Deltam$.

Within a Fermi liquid description the quasiparticle contribution to the Raman response in the superconducting state is described by ~\cite{LeTacon,Devereaux1}:
\begin{equation}\label{eq:response}
\chi^{''}_{\BAN,\BN}(\Omega)=
\frac{2\pi N_{F}}{\Omega}\left\langle \gamma^2_{\BAN,\BN}(\phi)\,(Z\Lambda(\phi))^2
\frac{\Delta(\phi)^2}{\sqrt{(\Omega)^2-4\Delta(\phi)^2}}\right\rangle_{FS}
\end{equation}

The angular average over the Fermi surface is denoted $\langle(\cdots)\rangle_{FS}$. $N_{F}$ is the density of states at the Fermi level, $\gamma_ {\BAN,\BN}$ are the Raman vertices which read $\gamma_{\BAN}(\phi)=\gamma^{0}_{\BAN}\cos 2\phi$ and
$\gamma_{\BN}(\phi)=\gamma^{0}_{\BN}\sin 2\phi$, respectively. $\Delta(\phi)^2/\sqrt{\Omega^2-4\Delta(\phi)^2}$ is a BCS coherence factor. Eq.~(9) can be easily deduced from Eq.~(7), (see Methods).

The function $Z(\phi)$ is the spectral weight of the Bogoliubov quasiparticles, while $\Lambda(\phi)$ is a Fermi liquid parameter associated with the coupling of these quasiparticles to the electromagnetic field. Expression (\ref{eq:response}) differs from simple BCS theory for a non interacting Fermi liquid at $T=0$ (see eq.5) by the presence of these quasiparticle renormalizations, which importantly only enter through the product $Z\Lambda(\phi)$.

In the following, we will show that the angular dependence of this quasiparticle renormalization plays a key role in accounting for the experimental observations.

Let us first consider the \BAN geometry the Raman vertex $\gamma_ {\BAN}(\phi)$ is peaked at the antinode $\phi=0$ which dominates the \BAN response, resulting in a pair-breaking coherence peak at $\hbar\Omega_{\BAN}=2\Deltam$ due to the singularity of the BCS coherence factor. The weight of this peak is directly proportional to the antinodal quasiparticle renormalization $(\ZAN)^2=(Z\Lambda)^2(\phi=0)$. Hence, the fact that the \BAN coherence peak looses intensity at low doping (and even disappears altogether at low doping) strongly suggests that $\ZAN$ decreases rapidly as doping is reduced. This has been first suggested in earlier Raman studies \cite{Chen}.

In the \BN geometry, the situation is more subtle because the Raman vertex is largest at the nodes, where the gap function (and hence the BCS coherence factor) vanishes. This is illustrated in fig.~11, pannels (AI-II).

\begin{figure}[!ht]
\begin{center}
\includegraphics[width=10cm]{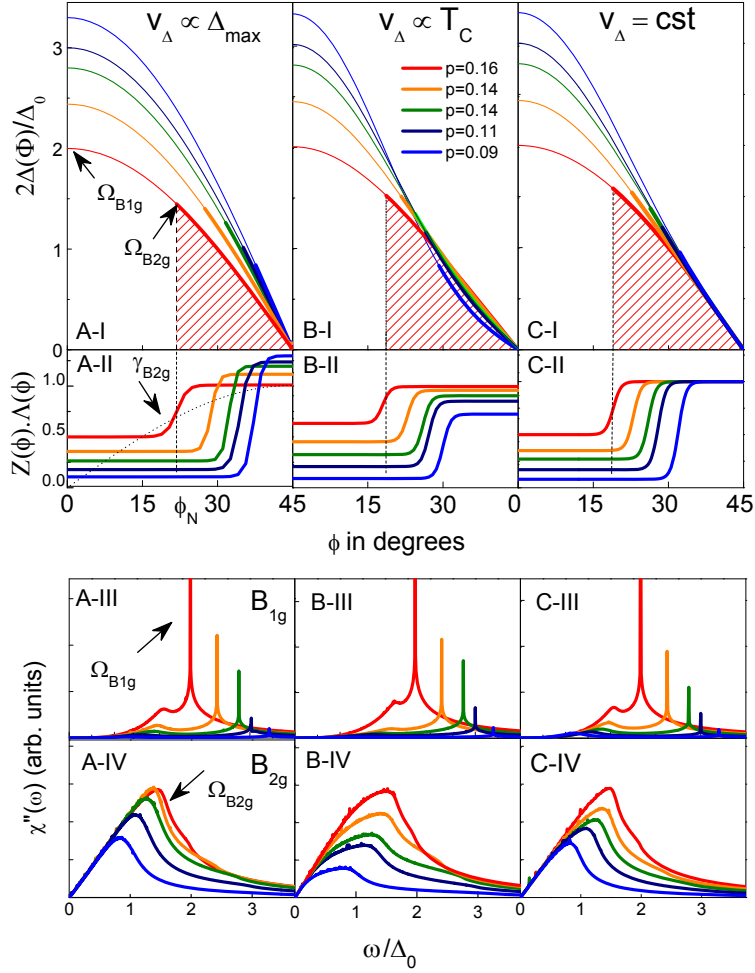}
\end{center}\vspace{-7mm}
\caption{Three scenarios for underdoped cuprates \cite{Blanc2}.
A.I-C.I: Doping evolution of the superconducting gap in the three scenarios (A-C).
A.II-C.II: Angular dependences of the quasiparticle spectral weights $Z\Lambda(\phi)$
as a function of doping level for each scenario. The angular dependence of the $B_{2g}$ Raman vertex
is shown in dotted line (see AII).
A.III-C.III, A.IV-C.IV: Calculated Raman spectra for each scenario in $B_{1g}$ and $B_{2g}$ geometries. The locations of the $B_{1g}$ and $B_{2g}$ peaks are respectively controlled by the gap energy at the antinodes $\Phi=0$ and at the angle $\Phi_{N}$.
Note that the low energy slope of the $B_{2g}$ Raman response is constant with doping as observed
experimentally \cite{Blanc}.}
\label{fig11}
\end{figure}

As a result, the energy of the coherence peak depends sensitively on the angular dependence of the quasiparticle renormalization $Z\Lambda(\phi)$. If the latter is approximately constant along the Fermi surface, then the energy of the \BN peak is determined solely by the angular extension of the Raman vertex $\gamma_{\BN}(\phi)$.

In contrast, let us consider a $Z\Lambda(\phi)$ which varies significantly from a larger value \ZN at the node to a small value \ZAN at the antinode, with a characteristic angular extension $\phi_N$ around the node smaller than the intrinsic width of the Raman vertex $\gamma_{\BN}(\phi)$. Then, it is $\phi_N$ itself which controls the position of the \BN peak: $\hbar\Omega_{\BN}=2\Delta(\phi_N)$.

As shown below, this explains the origin of the differentiation between the two energy scales in the underdoped regime.

To proceed further in the simplest possible way, we consider a simple crenel-like shape for $Z\Lambda(\phi)$, varying rapidly from $\ZN$ for $\phi_N<\phi<\pi/4$ to $\ZAN<\ZN$ for $0<\phi<\phi_N$ (see fig.~11, AII-CII).

Furthermore, we adopt the often-used~\cite{Mesot} parametrization of the gap function. $\Delta(\phi)=\Deltam\left[B\cos 2\phi + (1-B)\cos 6\phi\right]$, consistent with d-wave symmetry where the nodal slope of the gap $v_\Delta\equiv\partial\Delta/\partial\phi|_{\phi=\pi/4}=2(4B-3)\Deltam$ does not necessarily
track $\Deltam$. We thus have 5 parameters: $\Deltam$, \vD (or $B$), \ZAN, \ZN and the angular extension \phiN. These parameters are determined by attempting a fit to our spectra, obeying the following constraints: (i) the maximum gap \Deltam is determined from the measured energy of the \BAN peak according to $2\Deltam=\hbar\Omega_{\BAN}$; (ii)the antinodal quasiparticle renormalization \ZAN is determined such as to reproduce the intensity of the \BAN coherence peak; (iii) the angular extension \phiN is determined from the energy of the nodal coherence peak. Throughout the underdoped regime, this amounts to
$2\Delta(\phi_N)=\hbar\Omega_{\BN}$ as discussed above and finally the nodal renormalization \ZN is constrained to insure that the ratio
$(\ZN)^2/\vD$ does not change as a function of doping level, at least in the range $0.1 < p < 0.16$. This has been observed experimentally in Ref.~\cite{Blanc} and reported in fig.~7d.

This ratio controls the low-frequency behavior of the \BN Raman response. We assume here that the density of states $N_F$ (associated with the Fermi velocity perpendicular to the Fermi surface) does not depend sensitively on doping level in this range.

These 4 constraints leave one parameter undetermined, which can be taken as the deviation of the gap function from a pure $\cos k_x-\cos k_y$ form, as measured by the ratio $\vD/(2\Deltam)=4B-3$ of the nodal velocity to the maximum gap.

We will thus consider three possible scenarios:(A) Pure \cxcy gap: $\vD=\Deltam$ ($B=1$). This corresponds to a superconducting gap
involving a single characteristic energy, which increases as the doping level is reduced. (B) \vD tracks the critical temperature $T_c$. In this case, the gap function is truly characterized by two scales varying in opposite manner as the doping level is reduced. (C) \vD remains constant as a function of doping. This is also a two-scale superconducting gap scenario, although with a milder variation of \vD.

In figure~11, are displayed our fits of the \BAN and \BN Raman spectra in the framework of this simple theoretical analysis, following each of the three scenarios (A-C) above.

We observe that the main aspects of the experimental spectra, and most importantly the existence of two energy scales \OmAN, \OmN varying in opposite manners as a function of doping, can be reproduced within any of the three scenarios.

A common feature between all three scenarios is that the quasiparticle renormalization function $Z\Lambda(\phi)$ varies significantly along the Fermi surface. Quasiparticles have a large spectral \ZN only on a restricted region around the nodes, defined by $\phi_N$, corresponding to a fraction $\arc\equiv (\pi/4-\phi_N)/(\pi/4)$ of the Fermi surface.

While $\Deltam$ increases with falling doping, $\Delta(\phi_N)$ decreases because of the rapid contraction of the coherent fraction $\arc$, leading to the opposite doping dependence of the two scales, as illustrated on fig.~12-a.

\begin{figure}[!ht]
\begin{center}
\includegraphics[width=14cm]{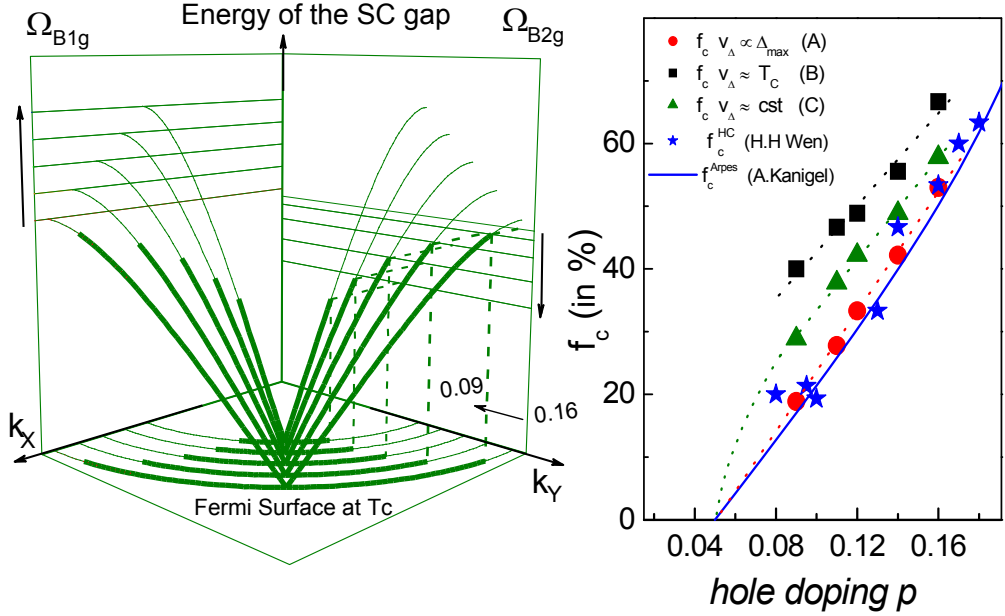}
\end{center}\vspace{-7mm}
\caption{(color online) Fermi surface coherent fraction and a single d-wave superconducting gap \cite{Blanc2}. (a) Scenario A where coherent Bogoliubov quasiparticles are partially suppressed on restricted parts of the Fermi surface and the superconducting gap has a single d-wave shape. (b) Doping evolution of the coherent fraction of the Fermi surface ($\arc$) in the three scenarios (A-C). The $\arc^{ARPES}$ curve is deduced from $L_{arc}/L_{full}(T_{c})=1-0.70\frac{T_{c}}{T^{*}}$
(see Ref.[\onlinecite{Kanigel2}]), expressed as a function of the doping level. $\arc^{HC}$ is extracted from the zero temperature specific heat coefficient in the normal state $\gamma_{n}(0)$ \cite{HHWen,Matsuzaki}.}
\label{fig12}
\end{figure}

We note that linearizing the gap function in the coherent region is a reasonable approximation for the A and C scenarios, leading to the relation $\hbar\OmN=\frac{\pi}{4}\arc\vD\propto k_BT_c$ which links the nodal (\BN) energy scale (proportional to $T_c$), the nodal velocity and the coherent fraction. \cite{Note2}. This approximation is not valid for the B scenario because it leads to $\arc$ as a constant in contradiction with this scenario. 

It is clear that having uniformly coherent Bogoliubov quasiparticles along the Fermi surface (corresponding to a constant $Z\Lambda(\phi)$) is inconsistent with our data, especially in view of the rapid suppression of the \BAN coherence peak and the corresponding decrease of \ZAN.

Although the above features are common to all three scenarios, there are two key differences between them. The first one is qualitative:
in scenario A  (a single gap $\vD\propto\Deltam$) the nodal renormalization factor  \ZN increases as doping level is reduced, while it {\it decreases} for scenario B ($\vD\propto T_c$) and stays constant for scenario C ($\vD\propto \rm{const.}$). 
The second, quantitative, difference is the rate at which the coherent fraction of the Fermi surface \arc decreases with falling doping, being largest for scenario A and smallest for B (Fig.~12-b).

Clearly, highly accurate spectroscopy measurements in the nodal region aiming directly at the determination of \vD or \ZN would discriminate between these three scenarios. Such measurements are, unfortunately, notoriously difficult and a consensus has not been yet reached.

A determination of the coherent fraction $\arc^{HC}$ has been reported from heat-capacity (HC) measurements~\cite{HHWen,Matsuzaki} and reproduced on Fig.~12-b. It was also reported from ARPES ~\cite{Kanigel,Kanigel2} in the normal state that the Fermi arcs shrink upon cooling as $\sim T/T^*$. The doping evolution of coherent fraction $\arc^{ARPES}$ at $T_c$ is displayed in Fig.12-b. 
Remarkably, we find that there is a good quantitative agreement between the doping dependence of \arc reported from HC and ARPES and our
determination from Raman within scenario A (a single gap scale $\vD\propto\Deltam$), which appears to be favored by this comparison.

Although this quantitative agreement should perhaps not be overemphasized in view of the uncertainties associated with each of the experimental probes, we conclude that this single-gap scenario (A) stands out as the most likely possibility.

We note that this interpretation reconciles the distinct doping dependence of the two energy scales with the thermal conductivity measurements of underdoped samples.  Thermal conductivity measurements interpreted within the clean limit and a Fermi velocity almost constant (in the doping range $p\sim 0.1-0.2$) show that $\vD\propto\Deltam$ \cite{Sutherland,Hawthorn}. 

We can further note by combining the ratios $(\ZN)^2/\vD \sim \rm{const.}$ (deduced from Raman and penetration depth measurements \cite{Blanc,Panagopoulos}) and $(\ZN)/\vD \sim p$ (deduced from heat capacity measurements under magnetic field \cite{HHWen}) that we get: $(\ZN)\sim 1/p$ and $\vD \sim 1/p^2$. Both $(\ZN)$ and $\vD$ increase with underdoping as expected within the single gap scenario (A). 
Here we have assumed that the nodal quasiparticles renormalization, $\ZN$,  takes roughly the same value for Raman, penetration depth and heat capacity measurements.

With a single-scale superconducting gap, the relation between the critical temperature (or $\OmN$) and the coherent fraction reads: $k_BT_c \propto \arc\Deltam$ (10). This is consistent with previous investigations \cite{Ding,HHWen} and more recent ones \cite{Yazdani,Ideta}. This relation carries a simple physical meaning, namely that it is the suppressed coherence of the quasiparticles that sets the value of $T_c$, while $\Deltam$ increases with falling doping. This relation differs from the standard BCS theory. Crucially, $T_c$ in cuprates depend on a prefactor, \arc which is doping dependent.  

We can also pointed out that scenario (A) is consistent with the Uemura relation (valid in the underdoped regime) and Homes'law  \cite{Uemura,Homes}. Indeed, $\rho_{S}\propto T_{c}$ and $\rho_{S}\propto \sigma_{dc}\Deltam$ (valid in the dirty limit) lead to $T_{c}\propto \sigma_{dc}\Deltam$ (11). $\rho_{S}$ and $\sigma_{dc}$ are respectively the superfluid density and dc conductivity.  By combining eq.(10) and (11) we obtain $\arc \propto \sigma_{dc}$ which makes sense since current flows on the fraction of coherent Fermi surface.

Finally, our interpretation is also in agreement with previous observations on Giaver and Andreev Saint-James (ASJ) tunneling experiments which pointed out the existence of two distinct energy scales in superconducting state of underdoped cuprates \cite{Deutscher} . The high energy scale was assigned to the single particle exictation energy. This is the energy of the first excited state required to break a Cooper pair in Giaver tunneling experiment \cite{Giaver}. This corresponds to the Raman \BAN scale associated to the pair breaking peak energy. The low energy scale was assigned to the energy range over which Cooper pairs can flow in the ASJ tunnelling. It is directly related to the Raman \BN scale since, this last one, is controlled by the fraction of coherent Fermi surface \arc around the nodes where supercurrents flow.

\section*{THE PSEUDO GAP AND LOSS OF ANTI-NODAL QUASIPARTCLES IN UNDERDOPED CUPRATES}

\subsection*{Raman experimental observation of the pseudo gap}

The exploration of the superconducting state of underdoped cuprates reveals that coherent Bogoliubov quasiparticles are reduced on restricted regions of the Fermi surface around the antinodes. This manifests itself by a strong decrease of the coherent Cooper pairs density at the antinodes while it is still sizeable around the nodes. We have defined a fraction of coherent Fermi surface around the nodes where the d-wave superconducting gap developps. The loss of quasiparticles spectral weight is then responsible for the strong dichotomy in the quasiparticles dynamics between the antinodal and the nodal regions and the emergence of two energy scales in the superconducting state.  

The loss of antinodal Bogoliubov quasiparticles in underdoped regime below $T_c$ is concomitant with a strong depletion of the \BAN electronic continuum in the normal state as the temperature is decreasing down to a temperature just above $T_c$.  This can be seen in fig.~13 for an underdoped $Bi-2212$ sample ($p\approx0.12$ and $T_c=75~K$). As the sample is cooled down from $T=250~K$ to $T=90~K$ the low energy electronic background level drecreases. Similar observations have been reported in earlier works \cite{Gallais,Venturini1}.

\begin{figure}[!ht]
\begin{center}
\includegraphics[width=8cm]{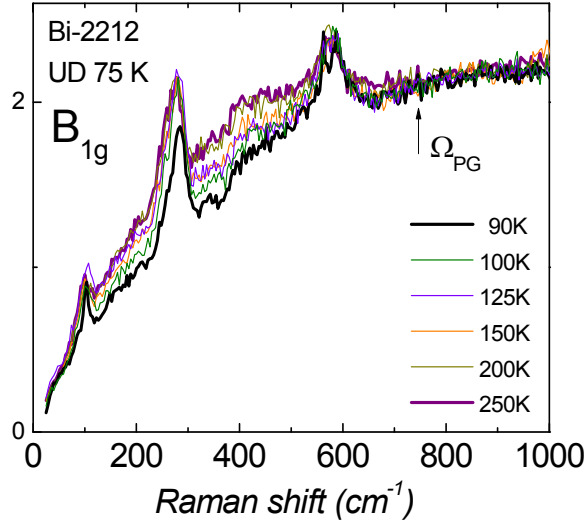}
\end{center}\vspace{-8mm}
\caption{Raman responses of underdoped $Bi-2212$ single crystals in $B_{1g}$(antinodal) as a function of temperature above $T_c$. A strong depletion of the electronic continuum is observed between $250~K$ and $90~K$ at low energy. The onset energy of the depletion is quoted $\Omega_{PG}$.}
\label{fig13}
\end{figure}

Such a coincidence leads us to wonder if the loss of quasiparticle spectral weight on restricted parts of the Fermi surface that we have revealed in the superconducting state of underdoped cuprates (see previous section) persists above $T_c$ and more generally is a salient feature of the underdoped cuprates physics.

In order to address this question we have performed Raman measurements on a large range of temperatures below and above $T_c$.
In fig.~14-a is displayed the temperature dependence of the Raman response of an underdoped $Bi-2212$ single crystal. For each temperature, the Raman response has been subtracted from the one measured at $T=250K$ and it has been reported on fig.~14-b.

\begin{figure}[!ht]
\begin{center}
\includegraphics[width=12cm]{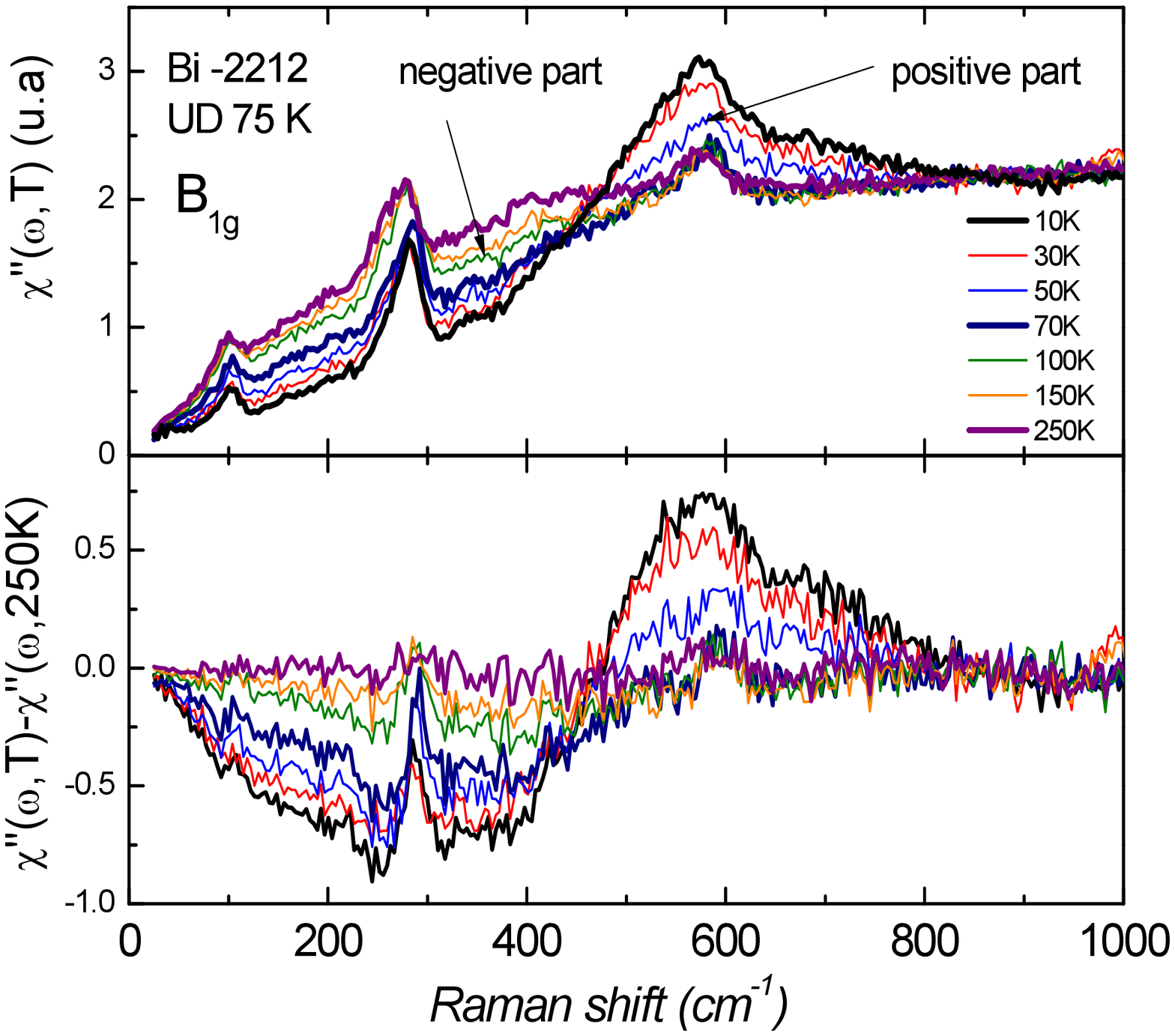}
\end{center}\vspace{-8mm}
\caption{
(a) Temperature dependence of the  $B_{1g}$ (antinodal) Raman spectra, (b) Subtraction of the $B_{1g}$ Raman from the one measured at $T=250~K$. We clearly distinguish a low energy negative part and a high energy positive part. The positive part disapears around $T_c$ while the negative one persists well above $T_c$.} 
\label{fig14}
\end{figure}

The subtracted Raman responses can then be decomposed in two parts: a positive and a negative one. The positive part (at high energy above $400~cm^{-1}$) developps in the superconducting state up to $T_{c}$ and corresponds to the coherent pairs breaking peak already discussed in the previous section. The negative part (below $400~cm^{-1}$) corresponds to the depletion of the low energy electronic continuum which is filled up when temperature increases. It persists well above $T_c$ and only disapears in the normal state around 200~K. The persistence of a low energy negative part well above $T_c$ is interpreted by us  as the experimental signature of the pseudogap in ERS. The lowest temperature from which the negative part is no more temperature dependent will set the pseudogap temperature $T*$. Here $T*$ is close to 200~K.

\begin{figure}[!ht]
\begin{center}
\includegraphics[width=15cm]{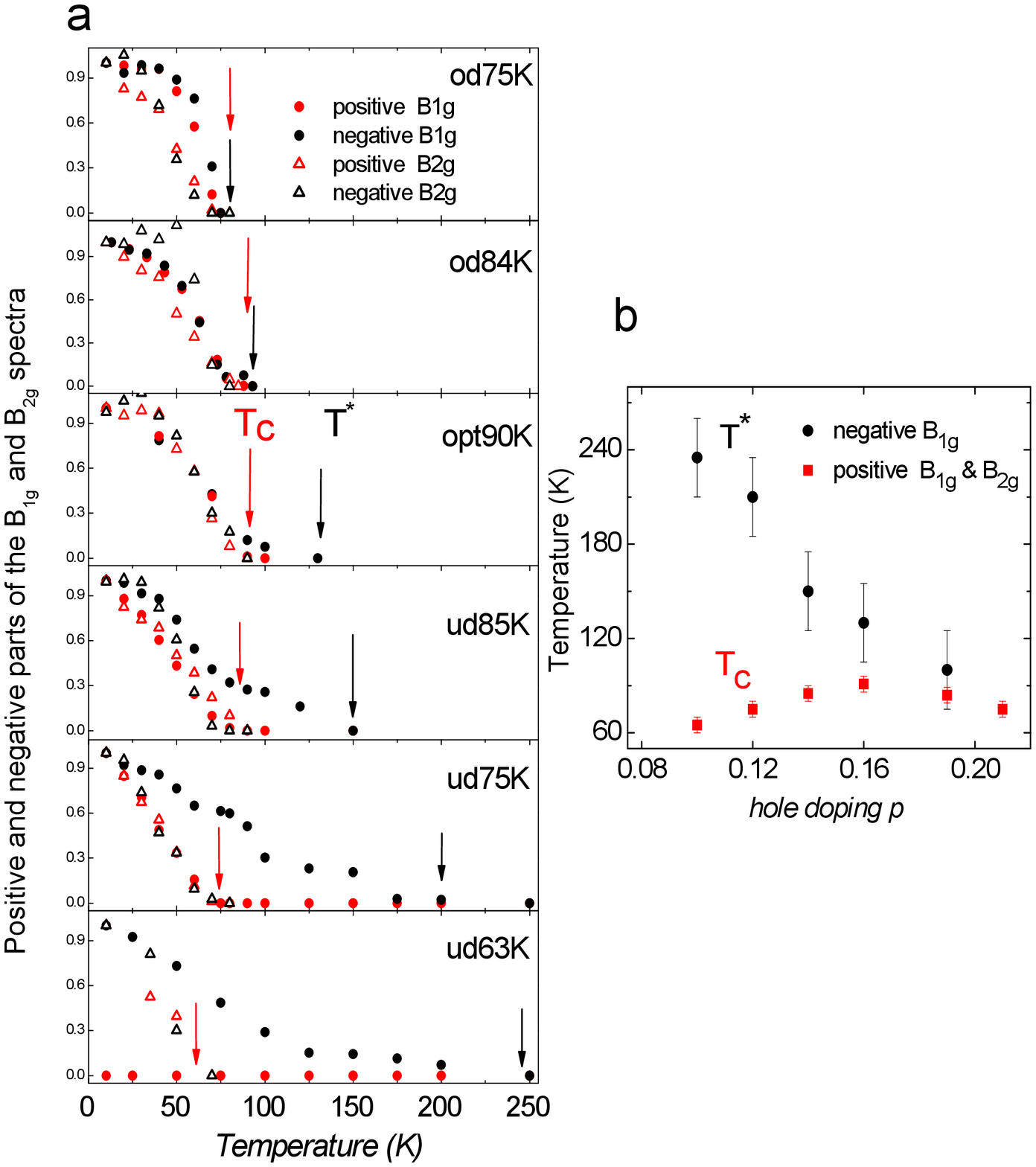}
\end{center}\vspace{-8mm}
\caption{(a)Temperature dependences of the positive and negative parts obtained from the subtraction of the \BAN and \BN Raman spectra of $Bi-2212$ for various doping levels. The positive and negative parts have been normalized to their values at $T=10~K$. The filled and open symbols refer to the \BAN and \BN geometries respectively. (b) Doping dependences of $T_c$ and $T*$ extracted from the positive parts of the \BAN and \BN Raman spectra and the negative part of the \BAN Raman spectra respectively.} 
\label{fig15}
\end{figure}

In fig.~15-a is displayed the temperature dependence of the positive and negative parts obtained from the subtracted \BAN (Anti-Nodal) and \BN (Nodal) Raman spectra of $Bi-2212$. The \BN Raman spectra will be shown in the next section. For each doping level, the \BAN and \BN Raman spectra have been subtracted by the lowest temperature from which no more depletion is detected. $T_c$ is indicated by a gray (red) arrow. We observe that the negative parts of the subtracted \BAN Raman spectra (black dot) persist well above $T_c$ in particular for low doping levels. Lower is the doping level higher is the temperature from which the depletion disappears. This gives us an estimate of $T*$ (marked by a black arrow). In contrast the \BN negative parts (related to the nodal region) (open black triangle) disappears around $T_c$. This shows that the pseudo gap is predominent in the antinodal region and becomes sizeable below the optimal doping level. This is consistent with ARPES data \cite {Kanigel,Kanigel2,Norman}. We are then able to extract the pseudogap temperature $T*$ from our experimental data. $T*\approx 130~K$ for an optimal doping level (Opt90K), $T*\approx 150~K$ for an underdoped sample with a $T_{c}=85~K$, beyond $T*\approx 200~K$ for an underdoped sample with a $T_{c}=75~K$ and in between $T*\approx 200~K$ and 250~K for an underdoped sample of $T_{c}=65~K$. The positive parts of the \BAN and \BN Raman spectra (red dot and open triangle) both disappear at $T_c$ as expected since they correspond to coherent Cooper pairs (see previous section). The doping dependence of $T*$ and $T_c$ extracted from our data are plotted in fig.15-b and it is consistent with the values of $T*$ reported by other techniques \cite{Nakano}. 

In summary, our experimental findings show that the pseudogap manifests itself in the normal state just above $T_c$, as a strong depletion of low lying electronic excitations at the antinodes which are only restored well above $T_c$ at $T*$. Simultaneously, in the superconducting state the coherent Bogoliubov quasiparticles are strongly reduced at the antinodes with underdoping (see fig.~7). This leads us to conclude that the pseudogap suppresses quasiparticles around the antinodes above and below $T_c$ in the underdoped regime. The pseudogap is then "`harmful"' to the formation of Cooper pairs and acts as a ''foe'' of superconductivity in underdoped cuprates. 

The doping dependence of $T*$ is still a subject of intense debate, although a consensus emerges concerning the underdoped regime where $T*$ is reported to increase as the doping level is reduced. In the overdoped regime the situation is not yet claryfied: three schemes are in discussion.  (i) $T*$ merges with $T_c$, (ii) crosses $T_c$ or (iii) ends at the superconducting dome in the overdoped regime \cite{Norman2}.

Let's return to our experimental findings in the superconducting state (see previous sections). We have defined two distinct regions in the superconducting state of the cuprates. The first one (in the overdoped side of superconducting dome) corresponds to a single energy scale. The \BAN and \BN energy scales merge together and quasipartciles are well defined over the whole Fermi surface. On the contrary, in the underdoped side of the supercondcuting dome, we clearly detect two distinct energy scales inside the superconducting state. The \BAN and \BN energy depart from each other and result from a loss of quasiparticles around the antinodes. Such a dichotomy inside the superconducting state which manifests itself by a loss of quasiparticles in the underdoped part of the superconducting dome leads us to think that $T*$ has to cross the superconducting dome.

\subsection*{Interpretation of the pseudo-gap}

In the underdoped regime, the pseudogap develops around the antinodes and suppresses quasiparticles below $T*$ which are not restored below $T_c$. A loss of quasiparticles in the antinodal region below $T*$ should manifest distinctly in the Raman spectra according to the \BAN or \BN geometries. We expect a depletion of the  \BAN (AN) electronic continuum level over a large energy range which is filled up as the temperature is rised before disappearing above $T*$ when coherent quasiparticles are restored over the whole Fermi surface. 

On the other hand, we expect no continuum depletion at low energy in the \BN since the pseudogap vanishes in the nodal region. The low energy quasiparticle dynamic is then expected to have a standard Fermi liquid behaviour. This means that the low energy slope of the \BN electronic continuum is expected to be proportional to the quasiparticle lifetime according to a simple Drude like model \cite{Zawadovwski} and increases with cooling.  The pseudogap only manifests itself in the \BN spectrum at higher energy by an electronic continuum depletion which disappears above $T*$ and involves the loss of quasiparticles induced by the end of the pseudogap amplitude away from the antinodes. 
This is indeed what is experimentally observed on the temperature dependence of the \BAN  and \BN  Raman spectra of an underdoped $Bi-2212$ single crystal ($T_c = 75K$) and are reported in fig.~16-a and b.

\begin{figure}[!ht]
\begin{center}
\includegraphics[width=10cm]{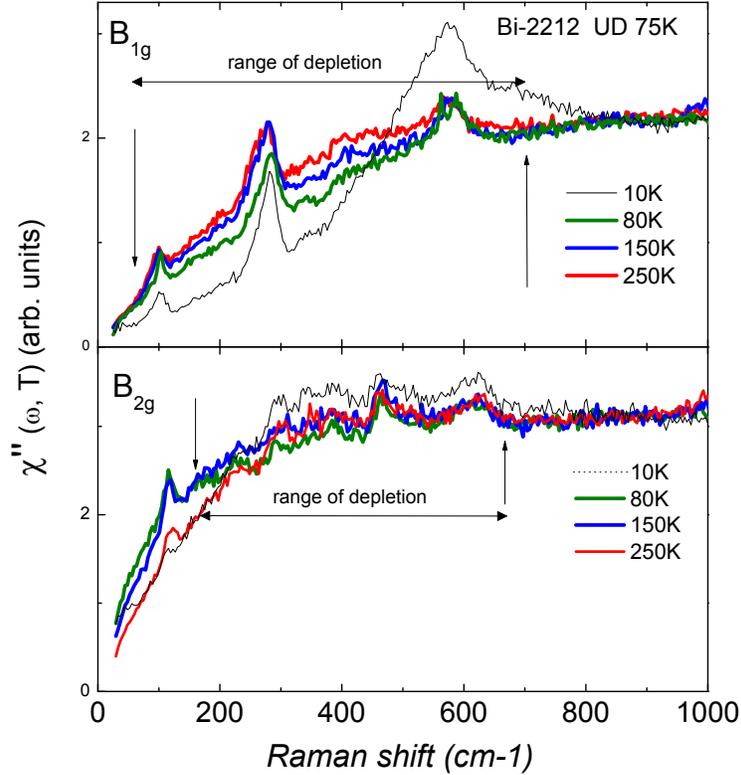}
\end{center}\vspace{-8mm}
\caption{Temperature dependences of the (a) \BAN and (b) \BN Raman responses of an underdoped $Bi-2212$ single crystal of $T_c = 75~K$. 
The low energy part of the \BAN electronic background displays a strong depletion at $80~K$ which is filled up at $T=250~K$. The \BN spectra  can be decomposed in two parts: a low energy part (below $180~cm^{-1}$) where the electronic background level decreases with heating and a high energy one (above $180~cm^{-1}$) where an electronic depletion is seen between etween 80~K and 150~K high energy part with a significant depletion at 80~K which dispears at 150~K and a low energy part which decreases as the temperature increases.} 
\label{fig16}
\end{figure}

In fig.~16-a, the \BAN electronic background exibits a strong depletion which is filled up as the temperature is rised from $80K$ to $250K$. This extends from $50$ to approximatively $700~cm^{-1}$. We can then notice that the onset of the depletion is localized close to the energy of the coherent peak. On the contrary, the \BN electronic background do not exhibits depletion below 180 $cm^{-1}$  and only a weak depletion between $80~K$ and $150~K$ which extends from 180 $cm^{-1}$ to approximatively 700 $cm^{-1}$. Such a weak depletion has been also observed in earlier works \cite{Nemetschek,Gallais}. We can notice that this weak depletion does not really affect the temperature dependence of the global \BN negative part reported in fig.~15.

Below 180 $cm^{-1}$ the low energy \BN background level decreases as the temperature increases in opposite way to the low energy \BAN electronic background level. This is a standard temperature dependence of the slope of an electronic Raman continuum free of a pseudogap. As the consequence, the low and high energy parts of the $B_{2g}$ Raman spectra exhibit two distinct temperature dependences in contrast to the $B_{1g}$ spectrum where the electronic background continuously increases as the temperature increases.

The experimental observations are different for the overdoped cuprates. In fig.~17-a and b are displayed the temperature dependence of the \BAN and \BN Raman spectra of an overdoped $Bi-2212$ single crytal ($T_{c}=84~K$). No clear depletion (which is filled up with heating) is observed in the \BAN and \BN electronic Raman continua. In \BN Raman spectra, solely, the decrease of the low energy slope of the electronic continuum is detected as the temperature is raised according to a simple Drude model.

\begin{figure}[!ht]
\begin{center}
\includegraphics[width=9cm]{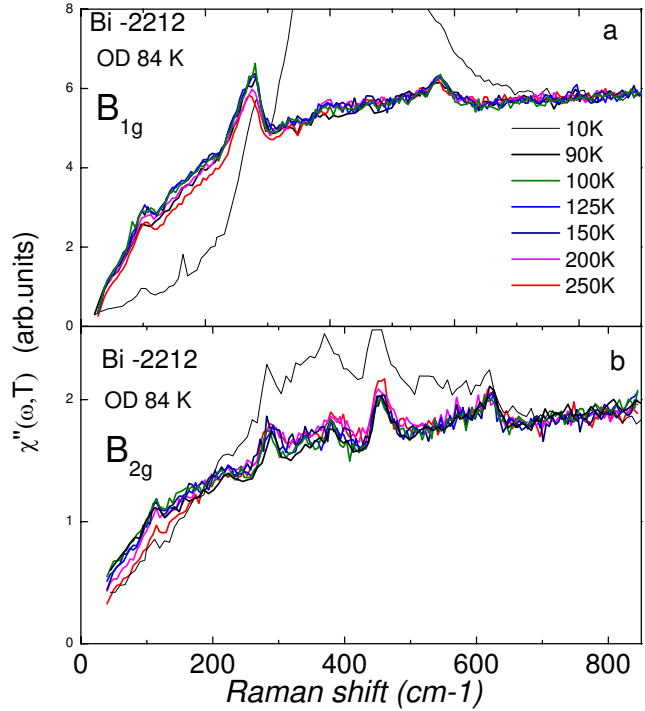}
\end{center}\vspace{-8mm}
\caption{Temperature dependences of the (a) \BAN and (b) \BN Raman responses of an overdoped Bi-2212 single crystal of $T_{c} = 75~K$. 
The low energy part of the \BN electronic background level decreases as the temperature is rised in the normal state. We observe the same trend for \BAN electronic background above $T = 150~K$.} 
\label{fig17}
\end{figure}

In summary our experimental findings in the normal state of underdoped cuprates have permitted us to detect the pseudogap in the electronic Raman responses and assign it to a partial suppression of coherent quasiparticles around the antinodes which are only restored above $T*$. 

This depletion which starts in the superconducting state of underdoped cuprates, is responsible for the strong decrease of the density of Cooper pairs in the antinodal region.

\section*{CONCLUSION}

As the Mott insulating is approaching, superconductivity is paradoxically confined in the nodal region where the superconducting gap amplitude is weak. k-space coherent Cooper pair islands are then formed in the nodal region while a strong decrease of coherent Bogoliubov quasiparticles in the antinodal region is experimentally observed. We have defined a fraction of coherent Fermi surface \arc upon which superconductivity developps around the nodal region. Below the optimal doping, \arc controls $T_c$ such as $k_BT_c \propto \arc\Deltam$. $T_c$ is decreasing while the superconducting gap amplitude $\Delta_{max}$ is increasing. This leads to two energy scales in the superconducting state of underdoped cuprates. Such a dichotomy in the quasiparticle dynamics persists in the normal state of underdoped cuprates up to $T*$ and is responsible for the emergence of the pseudogap phase which manifests itself as a loss of quasiparticles around the antinodes. Physics of the nodal quasiparticles is then predominant in underdoped cuprates. 

\subsection*{A tentative 3D phase diagram}

Let's now try to answer to the first question put in the introduction about a 3D phase diagram which involves both the energy and temperature phase diagrams. An attempt to depict such a phase diagram is shown in fig.~18.

\begin{figure}[!ht]
\begin{center}
\includegraphics*[width=14cm]{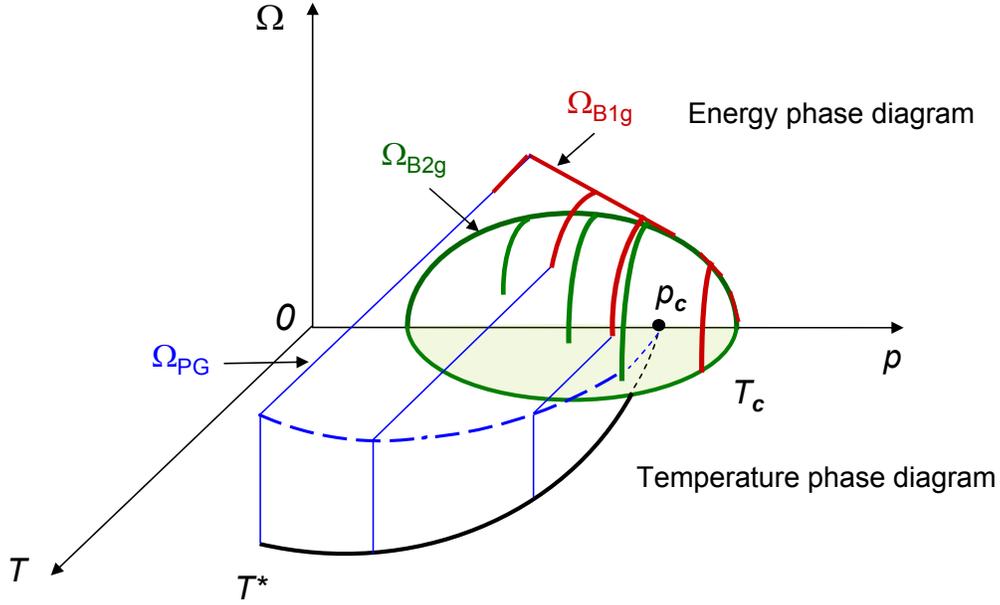}
\end{center}\vspace{-8mm}
\caption{3D Phase diagram: horizontal plane corresponds to the Temperature phase diagram versus doping level while the vertical plane to the Energy phase diagram versus doping level.}
\label{fig18}
\end{figure}

The energy phase diagram ($\Omega,p$) is plotted on the vertical plane. The \BAN and \BN energy scales (which correspond to the locations of the coherent peaks detected below $T_c$) are displayed. The upper \BAN energy scale is no more detected at low doping level while the lower \BN energy scale is still observable. The temperature phase diagram ($T,p$) is plotted in the horizontal plane. As suggested by our experimental findings, $T*$ cuts the superconducting dome in the overdoped side.  The $T_c$ and $T*$ curves are respectively in gray (green) and black.

The thin (blue) lines correspond to the onset energy of the \BAN depletion, $\Omega_{PG}$, detected below $T*$ and above $T_c$ (see fig.~13). Our preliminary results (see for an example figs.~14) seem to indicate that $\Omega_{PG}$ is roughly temperature independent and merges with the \BAN energy scale measured just below $T_c$. In fig.~18, the dashed (blue) line delimit the zone of the 3D phase diagram beyound which the \BAN depletion disappears and $\Omega_{PG}$ is no more detected.

In the strongly overdoped regime, above $p_{c}$, the $\Omega_{\BAN}$ and $\Omega_{\BN}$ energies exhibit the same temperature dependence and they both decrease as $T_c$ is reached. These observations have been reported in our previous studies \cite{Guyard01,Guyard02} and shown in figures 19-a and b. As the doping level is reduced, however the \BAN energy scale is no longer temperature dependent while the \BN energy scale is still temperature dependent and decreases (even slightly) as the temperature is increased up to $T_c$ (see fig.~19-a and b). This is sketched in fig.~18 by the appearance of two distinct branches for $\Omega_{\BN}$ and $\Omega_{\BAN}$ as the temperature increases.

Although more investigations are needed to track the doping dependence $\Omega_{PG}$ inside the superconducting dome, its extrapolation (drawn in fig.~18) seems to indicate that it rises from the $p_c$ doping level. $p_c$ is also the doping level from which the \BAN and \BN energy scales (in red and green respectively) depart form each other. Our interpretation is that $p_c$ is the starting doping level below which antinodal coherent quasiparticles become suppressed. Above $p_c$, superconductivity developps over the whole Fermi surface like in conventional superconductors. Below $p_c$ superconductivity is k-space confined and only developps on the fraction of coherent Fermi surface \arc around the nodes.

\begin{figure}[!ht]
\begin{center}
\includegraphics*[width=8cm]{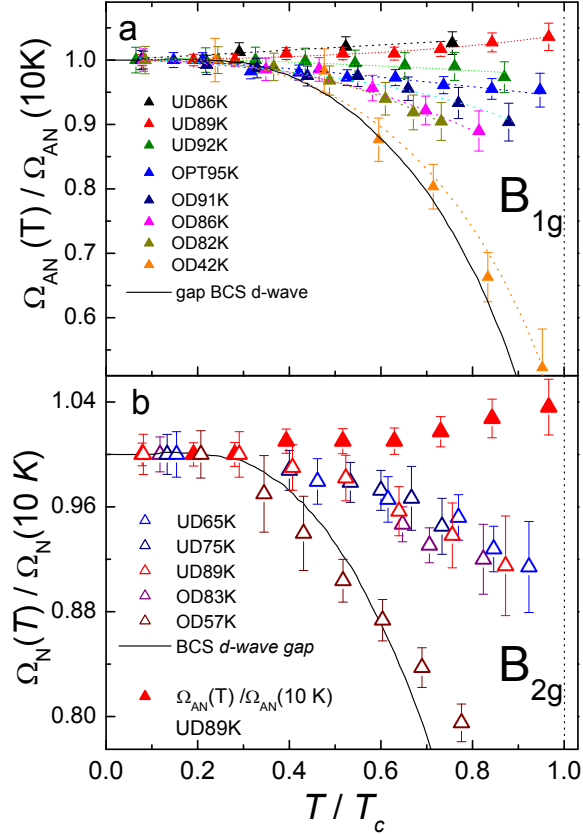}
\end{center}\vspace{-8mm}
\caption{Temperature dependences as a function of doping level of the $\BAN$ and $\BN$ peak energies \cite{Guyard01,Guyard02}.}
\label{fig19}
\end{figure}

\subsection*{How increasing Tc?}

Our experimental findings in the underdoped cuprates tell us that $T_c$ is limited by \arc such as $k_BT_c \propto \arc\Deltam$. The true challenge is then to increase \arc in the underdoped regime. In other words:  if we are able to restore a full Fermi surface in the underdoped regime of cuprates, we should increase $T_c$ in the cuprates. Unfortunately, loss of quasiparticles with underdoping around the antinodes acts against superconductivity and reduces $T_c$.

Why coherent quasiparticles at the antinodes are suppressed as the doping level is reduced? We do not know yet. Strong electron-electron interaction plays probably an important role in this mechanism.
We are then looking for cuprates with a normal resistivity as weak as possible above the optimal doping level (where both nodal and antinodal quasiparticles contribute to transport). We follow the simple idea that weak resistivity corresponds to weak electron-electron coupling. This contrasts with conventional superconductors where resistivity has to be chosen as large as possible to get a strong electron-phonon coupling. Here, the ``glue'' of pairing should be electron-electron interactions but if they are too strong, they reduce the coherent length and induce localization of Cooper pairs working against superconductivity.

\section*{METHODS} 

\subsection*{Details of the experimental procedure}
 
The ERS experiments have been carried out using a triple grating spectrometer (JY-T64000) equipped with a nitrogen cooled CCD detector.
ERS measurements give access to the dynamical structure factor, so the imaginary part of the Raman response function is obtained after correcting all the spectra for the Bose-Einstein factor. $\chi''(q,\omega,T)$ is also corrected for the spectral response of the spectrometer which involve mainly the efficency of the grattings and the CCD detector. In some special cases, for obtaining intrinsic Raman measurements from Bi-2212 crystals with various doping levels, we have placed an extremely high level of control of the crystal surface quality, the optical set up and the knowledge of optical constants for each crystal studied.  
 
ERS measurements on $Hg-1201$ single crystals have been performed with the red excitation line (1.9 eV) of a Kr+ laser to obtain the electronic  \BAN and \BN Raman response functions, the reason is that we have experimentally noticed that the Raman phonon activity is strongly reduced with the red line in comparison with the blue one~\cite{Kang,Gallais2,LeTacon2}. This gives us a direct view of the electronic response without invoking \textit{ad hoc} phonon subtraction procedures usually used in cuprate systems. In some cases, however, the green line has been used for probing the nodal region (\BN) due to the high efficiency of our detection in this energy range. Raman measurements on $Bi-2212$ single crystals have been carried out mostly with the green excitation line (2.4 eV).

\subsection*{Crystal growth}

The $Hg-1201$ single crystals have been grown by the flux method, and oxygen annealing have been performed in order to overdope crystals [\onlinecite{Bertinotti}]. The oxygen annealing is efficient within a few $\mu$m from the surface which is satisfactory for performing Raman scattering (the light penetration depth being of the order of the $100~nm$). The underdoped crystals are homogeneous as-grown single crystals. $Hg-1201$ is a quite ideal cuprate material for ERS measurements see ref. [\onlinecite{Guyard01,LeTacon}] for more details. It takes a pure tetragonal symmetry without any Cu-O chain contrary to $YBa_{2}Cu_2O_{7-\delta}$ (Y-123) or buckling which alters the unit cell of $Bi-2212$. We can then separately measure pure nodal and antinodal Raman responses, without mixing effects. Hg-$1201$ is made of one single $CuO_2$ layer which is a plane of symmetry in the unit cell. Raman active modes are therefore forbidden in the $CuO_2$ layer. This allows us to investigate the low energy electronic Raman spectrum without being hindered by extra phonons lines. 

The $Bi-2212$ single crystals were grown by using a floating zone method \cite{Gu}. 
$Bi-2212$ system can be easily cleaved providing large homogeneous surfaces ($\approx mm^{2}$). By using the same protocol as the one developped elsewhere \cite{Blanc}, we have obtained intrinsic Raman measurements to make reliable quantitative comprarisons between the Raman intensities of crystals with distinct doping level.

For both $Hg-1201$ and $Bi-2212$ systems, the doping value $p$ is inferred from $T_c$ using Presland and Tallon's equation: $1-T_{c}/T_{c}^{max} = 82.6 (p-0.16)^{2}$ \cite{Presland} and $T_{c}$ has been determined from magnetization susceptibility measurements for each doping level.   

\subsection*{Derivation of equation (8) from equation (5)}

When $T\rightarrow~0$, the Raman susceptibility becomes: 
\begin{eqnarray}
\chi^{,,}_{\mu}(\Omega) =\pi  \sum_{k}(\gamma^{\mu}_{k})^{2}\frac{\left|\Delta_k\right|^{2}}{E_{k}^{2}}\delta(\Omega-2E_{k})
\end{eqnarray}

with $E_{k}=\sqrt{\xi_{k}^{2}+ \Delta^{2}_{k}}$

We can first notice that (i) the integrant is predominant when $\xi_{k}$ goes to zero which means $\epsilon_{k}\approx\epsilon_{F}$ and (ii) $\delta(\Omega-2E_{k})$ is a function of $\xi_{k}$. This means

$\delta(f(\xi_{k}))=\sum_{i}\frac{1}{\left|f'(\xi^{i}_{k})\right|}\delta(\xi^{i}_{k}-\xi_{k})$.
Here $\xi^{i}_{k}$ are the zeros of $\xi_{k}$.

We consider a 2D Brillouin zone (convenient for cuprates) and we transform the sum over $k$ by an integration such as ``$dk$''is chosen perpendicular to the constant-energy line. We put:

$\frac{dk}{2\pi}=\frac{d\epsilon_{k}}{2\pi\left|\nabla\epsilon_{k}\right|}=N_{\bot}(\epsilon_{k})$.

We then obtain for a non interacting Fermi liquid:

\begin{eqnarray}
\chi^{,,}_{\mu}(\Omega) = N_{F}\int d\phi (\gamma^{\mu}(\phi))^{2}\frac{\left|\Delta(\phi)\right|^{2}}{\sqrt{\Omega^{2}-4\Delta^2(\phi)}}
\end{eqnarray}

where $N_{F}= N_{\bot}(\epsilon_{F})\frac{k_{F}}{2\pi}$

For an interacting Fermi liquid, we have to take into account the square of the quasiparticle renormalization factor $Z\Lambda(\phi)$ which then appears in the integrant of (eq.11) \cite{LeTacon}.

\section*{ACKNOWLEDGEMENTS}
We are grateful to  A.~Georges, Ph.~Bourges, J.~Carbotte, J.~C.~Campuzano, J.~Mesot, M.~Le~Tacon, G.~Kotliar, G.~Blumberg, L.~Taillefer, N.~Hussey, J.~Tallon, H.~H.~Wen, D.~Pavuna, J.~C.~Davis, A.~Yazdani, C.~Ciuti, R.~Lobo, M.~Civelli, A.~J.~Millis, P.~Coleman, P.~Hirschfeld, Ph.~Monod, F.~Rullier Albenque , C.~ C.~Homes and Z.~Tesanovic for very helpful discussions. The authors A. S., Y. G., M. C. and S. B. would like to thank support from french national agency for research (ANR), BLAN07-1-183876, GAPSUPRA.  Correspondences and requests for materials should be adressed to A.S.(alain.sacuto@univ-paris-diderot.fr).


\begin{thebibliography}{9}
\bibitem{Note} This analysis makes the assumption that for an interacting Fermi liquid (as it is the case for cuprates), the renormalized quasiparticles spectral weight which comes from the coupling between quasiparticles in interaction and the electromagnetic field follows the same doping dependence than the coherent Bogoliubov quasiparticles.
\end{thebibliography}
\end{document}